\documentclass[acmtog]{acmart}

\usepackage{booktabs} 
\usepackage{mathtools}

\usepackage{amsmath}
\DeclareMathOperator*{\argmax}{\arg\!\max}
\DeclareMathOperator*{\argmin}{\arg\!\min}
\usepackage{bbm}

\usepackage[ruled]{algorithm2e} 

\SetAlFnt{\small}
\SetAlCapFnt{\small}
\SetAlCapNameFnt{\small}
\SetAlCapHSkip{0pt}

\acmVolume{0}
\acmNumber{0}
\acmArticle{0}
\acmYear{2020}
\acmMonth{0}

\setcopyright{rightsretained}

\received{May 2020}

\begin{document}
\title{SceneGen: Generative Contextual Scene Augmentation using Scene Graph Priors}

\author{Mohammad Keshavarzi}
\orcid{0000-0003-2881-165X}
\affiliation{%
 \institution{University of California, Berkeley}
  \streetaddress{337 Cory Hall}
  \city{Berkeley}
  \state{California}
  \postcode{94720}
  \country{USA}}
\email{mkeshavarzi@berkeley.edu}

\author{Aakash Parikh}
\affiliation{%
 \institution{University of California, Berkeley}
  \city{Berkeley}
  \state{California}
  \postcode{94720}
  \country{USA}}
\email{aparikh98@berkeley.edu}

\author{Xiyu Zhai}
\affiliation{%
 \institution{University of California, Berkeley}
  \city{Berkeley}
  \state{California}
  \postcode{94720 (work done previous to joining Amazon)}
  \country{USA}}
\email{xiyu.z@berkeley.edu}

\author{Melody Mao}
\affiliation{%
 \institution{University of California, Berkeley}
  \city{Berkeley}
  \state{California}
  \postcode{94720}
  \country{USA}}
\email{melody_mao@berkeley.edu}

\author{Luisa Caldas}
\affiliation{%
 \institution{University of California, Berkeley}
  \streetaddress{XR Lab, Wurster Hall}
  \city{Berkeley}
  \state{California}
  \postcode{94720}
  \country{USA}}
\email{lcaldas@berkeley.edu}

\author{Allen Y. Yang}
\affiliation{%
 \institution{University of California, Berkeley}
  \streetaddress{337 Cory Hall}
  \city{Berkeley}
  \state{California}
  \postcode{94720}
  \country{USA}}
\email{allenyang@berkeley.edu}

\begin{abstract}
Spatial computing experiences are constrained by the real-world surroundings of the user. In such experiences, augmenting virtual objects to existing scenes require a contextual approach, where geometrical conflicts are avoided, and functional and plausible relationships to other objects are maintained in the target environment. Yet, due to the complexity and diversity of user environments, automatically calculating ideal positions of virtual content that is adaptive to the context of the scene is considered a challenging task. Motivated by this problem, in this paper we introduce SceneGen, a generative contextual augmentation framework that predicts virtual object positions and orientations within existing scenes. SceneGen takes a semantically segmented scene as input, and outputs positional and orientational probability maps for placing virtual content. We formulate a novel spatial Scene Graph representation, which encapsulates explicit topological properties between objects, object groups, and the room. We believe providing explicit and intuitive features plays an important role in informative content creation and user interaction of spatial computing settings, a quality that is not captured in implicit models. We use kernel density estimation (KDE) to build a multivariate conditional knowledge model trained using prior spatial Scene Graphs extracted from real-world 3D scanned data. To further capture orientational properties, we develop a fast pose annotation tool to extend current real-world datasets with orientational labels. Finally, to demonstrate our system in action, we develop an Augmented Reality application, in which objects can be contextually augmented in real-time.
\end{abstract}

%
%

\ccsdesc[500]{Computing methodologies~Mixed / augmented reality}
\ccsdesc[300]{Mathematics of computing~Kernel density estimators}
\ccsdesc[300]{Computing methodologies~Virtual reality}
%
%

\keywords{Augmented Reality, Scene Graphs, Scene Synthesis, Generative Modelling, Spatial Computing}
\begin{teaserfigure}
  \includegraphics[width=\textwidth]{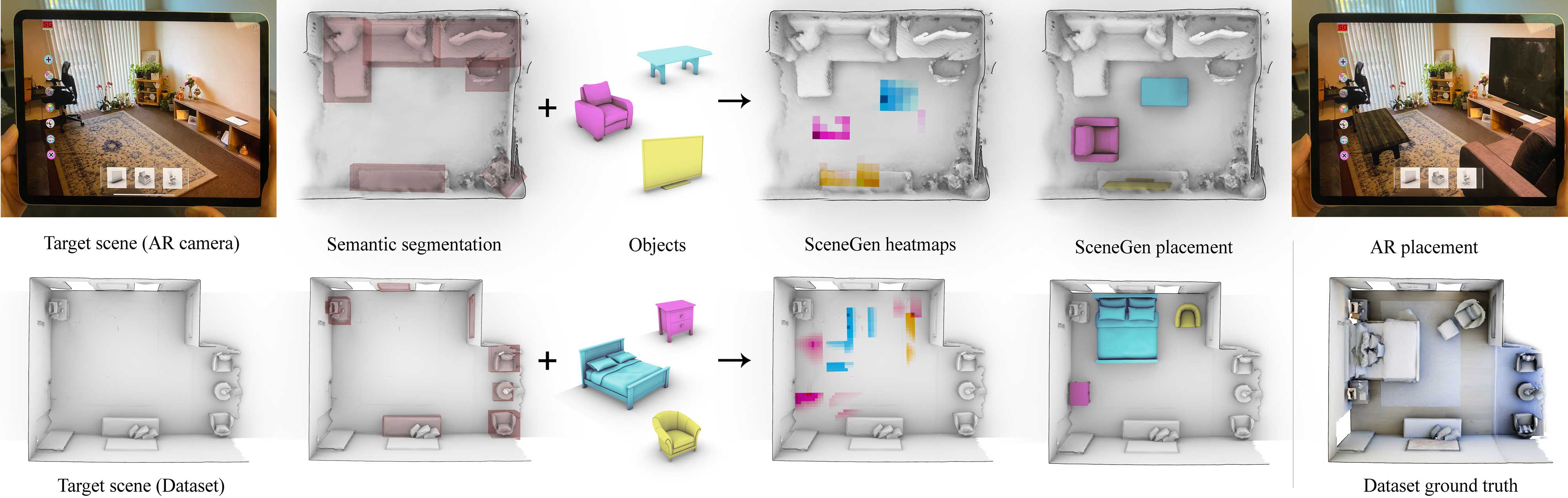}
  \caption{SceneGen is a framework to augment scenes with virtual objects using an explicit generative model to learn topological relationship from priors extracted from a real-world datasets. Primarily designed for spatial computing applications, SceneGen extracts features from rooms into a novel spatial Scene Graph representation and iteratively augments objects by sampling positions and orientations in the scene to create a probability map and predicts a viable contextual placement for the virtual object.}
  \label{fig:teaser}
\end{teaserfigure}

\maketitle{}

\section{Introduction}
Spatial Computing experiences such as \emph{augmented reality} (AR) and \emph{virtual reality} (VR) have formed a newly exciting market in today’s technological space. New applications and experiences are being launched daily across the categories of gaming, healthcare, design, education, and more. However, for all of the countless applications available, they are physically constrained by the geometry and semantics of the 3D user environment where existing furniture and building elements are present \cite{narang2018simulating, Razzaque2001} . Contrary to traditional 2D graphical user interface, where a flat rectangular region hosts digital content, 3D spatial computing environments are often occupied by physical obstacles that are diverse and often times non-convex. Therefore, how one can assess content placement in spatial computing experiences is highly dependent on the user’s target scene. 

However, since different users may reside in different spatial environments, which differ in dimensions, functions (rooms, workplace, garden, etc.), and open usable spaces, existing furniture and their arrangements are often unknown to the developers, making it very challenging to design a virtual experience that would adapt to all user’s environments. Therefore, contextual placement is currently addressed by asking users themselves to identify the usable spaces in their surrounding environments or manually positioning the augmented object(s) within the scene. Currently, virtual object placement in most AR experiences is limited to specific surfaces and locations, e.g., placing objects naively in front of the user with no scene understanding, or only using basic horizontal or vertical surface detection. These simplistic strategies can work to some extent for small virtual objects, but the methods break down for larger objects or complex scenes with multiple object augmentation requirements. This limitation is further elevated in remote multi-user interaction scenarios, where finding a common virtual ground physically accessible to all participants to augment their content becomes challenging. \cite{keshavarzi2020optimization}. Hence, such experiences automatically become less immersive once the users encounter implausible virtual object augmentation in their environments.

\begin{figure}
  \includegraphics[width=0.95\columnwidth]{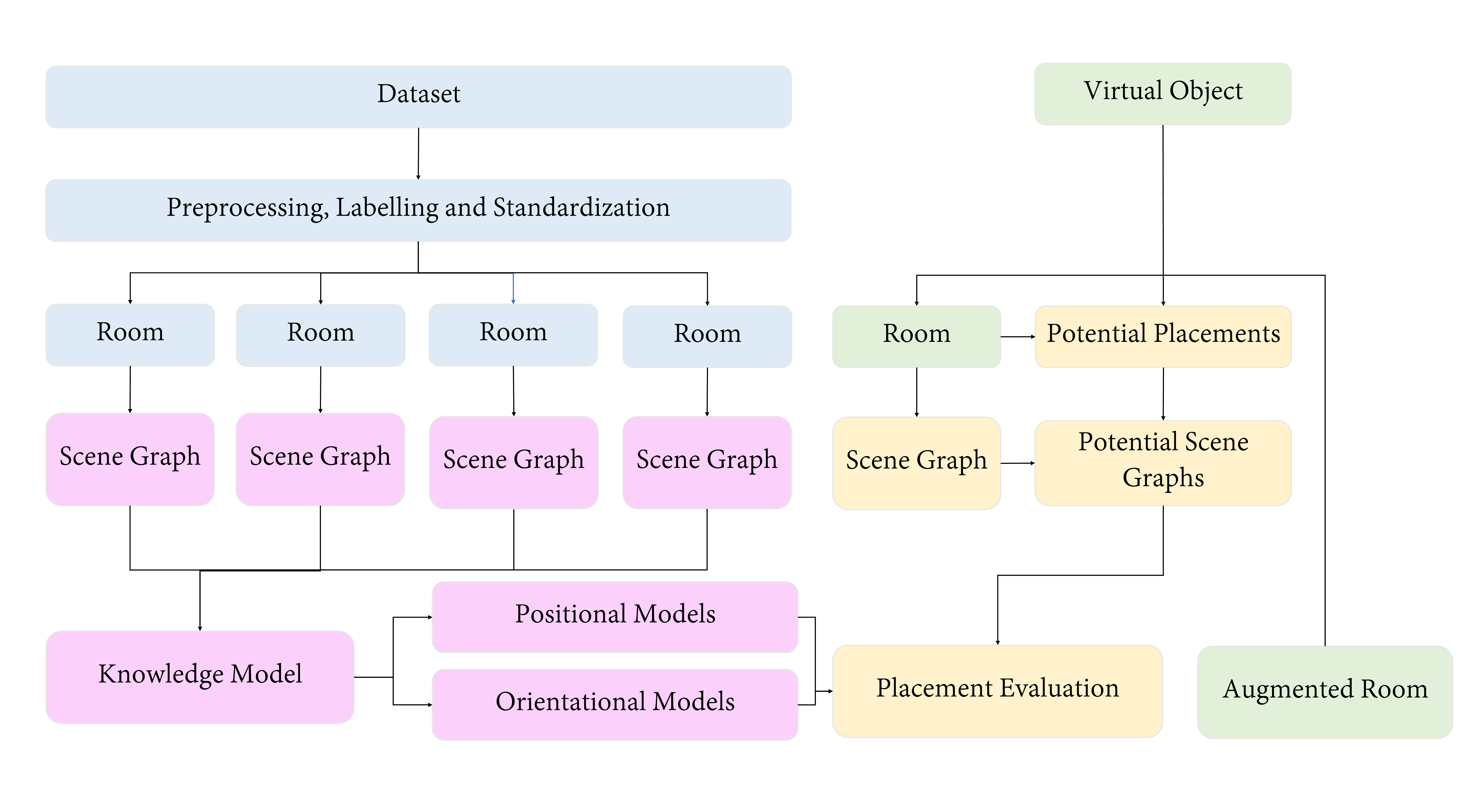}
  \caption[Workflow of SceneGen]{End-to-end workflow of SceneGen shows the four main modules of our framework to augment rooms with virtual objects. The left pipeline shows the training procedure including dataset processing (blue) and Knowledge Model creation (pink). The right pipeline shows the test time procedure of sampling and prediction (yellow) and the application (green).}~\label{fig:workflow}
\end{figure}

The task of adding objects to existing constructed scenes falls under the problem of \emph{constrained scene synthesis}. The work of \cite{kermani2016learning, ma2016action, li2019grains, Qi2016, ritchie2019fast, wang2019planit} are examples of such approach. However, there are currently two major challenges in the general literature which also create bottlenecks for virtual content augmentation in spatial computing experiences. First, current scanned 3D datasets publicly available are limited in size and diversity, and may not offer all the data required to capture topological properties of the rooms. For instance, \textit{pose},  the direction in which the object is facing, is a critical feature for understanding the orientational property of an object. Yet, such property is not clearly annotated for all objects in many large-scale real-world datasets such as SUN-RGBD and Matterport3D. Therefore, more recent research has adapted synthetic datasets, which can be used to extract higher-level information such as pose as they do not necessarily need to be manually annotated.

However, a critical drawback of synthetics datasets is that it cannot capture the natural transformation and topological properties of objects in real-world settings. Furniture in real-world settings are a product of gradual adoption of a space, contributing to the functionality of the room and surrounding items. Topological relationships between objects in real-world scenes typically exceed theoretical design assumptions of a architect, and instead capture contextual relationships from a living environment. Moreover, the limitations of the modeling software for synthetic datasets can also introduce unwanted biases to the generated scenes. The SUNCG \cite{song2016ssc} dataset, for instance, was built with Planner5D platform, an online tool which any user around the world can use. However, it comes with modeling limitations for generating rooms and furniture. Orientations are also snapped to right angles by default, which makes most scenes in the dataset Manhattan-like. More importantly, there is no indication if the design is complete or not, namely, a user may just start playing with the software and then leave at a random time, while the resulting arrangement is still captured as a legitimate  human-modeled arrangement in the dataset.

Second, recent models take advantage of implicit deep learning models and have shown promising results in synthesizing indoor scenes. Yet, their approach falls short for content developers to parameterize customized placement in relation to standard objects in the scene, and to generate custom spatial functionalities. One major limitation of these studies is that they do not have direct control over objects in the generated scene. For
example, authors of \cite{li2019grains} report they cannot specify object counts or constrain the scene to contain a subset of objects. Such limitations come from the implicit nature of such networks. Implicit models produce a black-box tool, which 
is difficult to comprehend should a end-user wishes to tweak its functions.
In cases where new objects are set to be placed, implicit structures may not provide abilities to manually define new object types, without providing . Moreover, using deep convolution networks require large datasets to train, a bottleneck that we have discussed above.

Motivated by these challenges, in this paper we introduce SceneGen, a generative contextual augmentation framework that provides probability maps for virtual object placements. Given a non-empty room already occupied by furniture, SceneGen provides a model-based solution to add new objects in functional placements and orientations. We also propose an interactive generative system to model the surrounding room. Contrary to the unintuitive implicit models, SceneGen is based on clear, logical object attributes and relationships. In light of the existing body of literature on semantic Scene Graphs, we leverage this approach to encapsulate the relevant object relationships for scene augmentation. Scene Graphs have already been in use for general scene generation tasks; they can also inform the intelligent placement of virtual objects in physical scenes.  

We use kernel density estimation (KDE) to build a multivariate conditional model to encapsulate explicit positioning and clustering information for object and room types. This information will allow our algorithm to determine likely locations to place a new object in a scene while satisfying their physical constraints. Object orientations are predicted using a probability distribution. From the calculated probabilities, we generate a score for each potential placement of the
new object, visualized as a heat map over the room. Our system is user-centric and ensures that the user understands the influence of data points and object attributes on the results. In addition, recent work has produced extensive scans of real-world environments. We use one such dataset, Matterport3D \cite{Chang2018}, in place of synthetic datasets such as SUNCG. As a trade-off, our real-world environment data are prone to messy object scans and non-Manhattan alignments.

Our contributions can be summarized as follows:
\begin{enumerate}
    \item We introduce a spatial Scene Graph representation which encapsulates positional and orientational relationships of a scene. Our proposed Scene Graph captures pairwise topology between objects, object groups, and the room.
    
    \item We develop a prediction model for object contextual augmentation in existing scenes.  We construct an explicit Knowledge Model which is trained from Scene Graph representations captured from real-world 3D scanned data. 
    
    \item To learn orientational relationships from real-world 3D scanned data, we have labeled the Matterport3D dataset with pose directions by human. To do so, we have developed an open-source labeling tool for fast pose labeling.
    
    \item We develop an Augmented Reality (AR) application that scans a user's room and generates a Scene Graph based on the existing objects. Using our model, we sample poses across the room to determine a probabilistic heat map of where the object can be placed. By placing objects in poses where the spatial relationships are likely, we are able to augment scenes that are realistic.
    
\end{enumerate}

We believe our proposed system can facilitate a wide variety of AR/VR applications. For example collaborative environments require placing one user’s objects into another user’s surroundings. More recently, adding virtual objects to scenes has been explored in online-shopping settings. This work can also apply to design industries, for example in generating 3D representations of example furniture placements. In addition, content creation of augmented and virtual reality experiences requires long hours of cross platform development on current applications, so our system will allow faster scene generation and content generation in AR/VR experiences.

\begin{figure}
  \includegraphics[width=1\columnwidth]{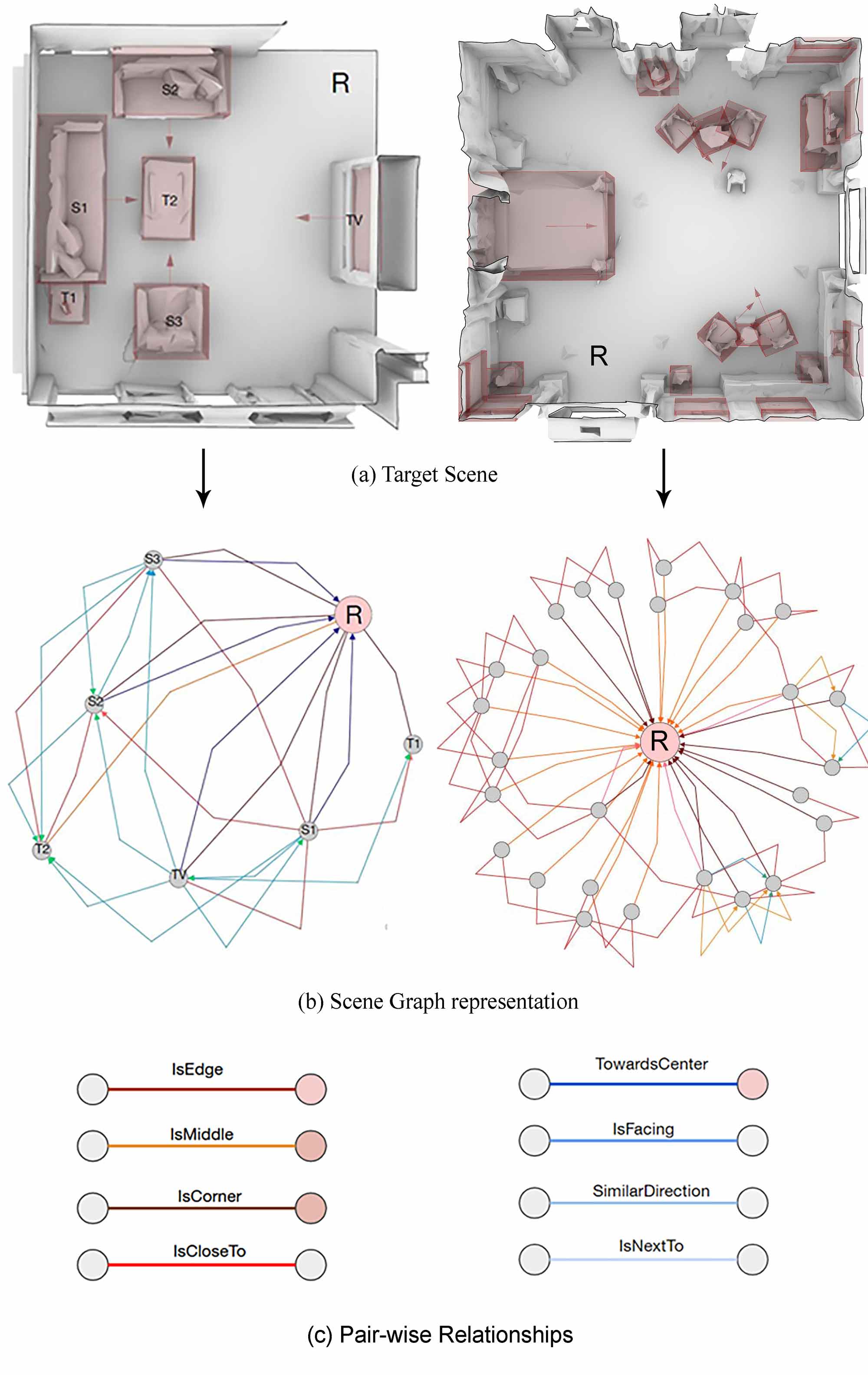}
  \caption[SceneGraph]{Our proposed Scene Graph representation is extracted from each scene capturing orientation and position based relationships between objects in a scene (pairwise) and between objects and the room itself. Visualization shows a subset of features for clarity.}~\label{fig:sceneGraph}
\end{figure}

\begin{figure*}
  \includegraphics[width=1.95\columnwidth]{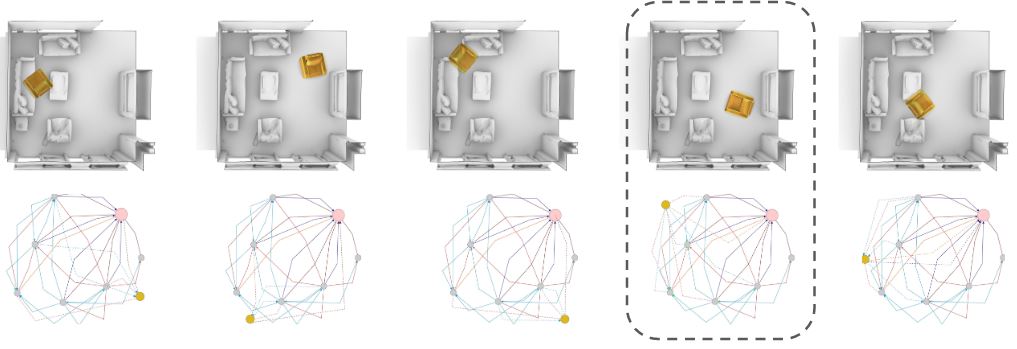}
  \caption[Graph Options]{Each placement choice for an object forms different topological relationships captured by the Scene Graphs. SceneGen evaluates the probability of these new relationships to create a probability map and recommend a placement.}~\label{fig:graphoptions}
\end{figure*}

Source code and pretrained models for our system can be found at our website after the review.

\section{Related Work}

\subsection{Scene Understanding}
Semantic Scene Graphs form one part of the overall task of scene understanding. Given visual input, as AR experiences generally would receive, one can tackle the tasks of 3D scene reconstruction and visual relationship detection. On the latter topic, a progression of papers attempted to encapsulate human "common-sense" knowledge in various ways: physical constraints and statistical priors \cite{silberman2012indoor}, physical constraints and stability reasoning \cite{jia20133d}, physics-based stability modeling \cite{zheng2015scene}, language priors \cite{lu2016visual}, and statistical modeling with deep learning \cite{dai2017detecting}. A similar approach was detailed in \cite{kim2012acquiring} for 3D reconstruction, taking advantage of the regularity and repetition of furniture arrangements in certain indoor spaces, e.g., office buildings. In \cite{xu20163d}, the authors proposed a technique that potentially could be well suited to AR applications, as it builds a 3D reconstruction of the scene through consecutive depth acquisitions, which could be taken incrementally as a user moves within their environment. Recent work has addressed problems like retrieving 3D layouts from 2D panoramic input \cite{sun2019horizonnet, kotadia2020indoornet} or floorplan sketches \cite{keshavarzi2020sketchopt}, building scenes from 3D point clouds \cite{pittaluga2019revealing, shi2019hierarchy}, and 3D plane reconstruction from a single image \cite{yu2019single, liu2019planercnn}. One can consult a recent overview of the topic in \cite{liu2019visual}. Our approach leverages this work on scene understanding, because our model operates on the assumption that we already have locations and bounding boxes of the existing objects in scene.

\subsection{Semantic Scene Graphs}
Semantic Scene Graphs have been applied to various tasks in the past, including image retrieval \cite{johnson2015image}, visual question answering \cite{teney2017graph}, image caption generation \cite{yao2018exploring}, and more. The past research can be divided into two approaches: (1) separate stages of object detection and graph inference, and (2) joint inference of object classes and graph relationships. Papers that followed the first approach often leverage existing object detection networks \cite{ren2015faster,li2017scene,zellers2018neural,yao2018exploring,chen2019knowledge}. Similarly to other scene understanding tasks, many methods also involved learning prior knowledge of common scene structures in order to apply them to new scenes, such as physical constraints from stability reasoning \cite{yang2017support} or frequency priors represented as recurring scene motifs \cite{zellers2018neural}. Most methods were benchmarked based on the Visual Genome dataset \cite{krishna2017visual}. However, recent studies found this dataset to have an uneven distribution of examples across its data space. In response, researchers in \cite{gu2019scene} and \cite{chen2019knowledge} proposed new networks to draw from an external knowledge base and to utilize statistical correlations between objects and relationships, respectively. Our work focuses on the task of construction and utilization of the semantic Scene Graph. As in \cite{zellers2018neural,chen2019knowledge}, we also use statistical relationships and dataset priors; but unlike these papers, we use a mathematical model rather than deep learning. Because our approach is based on a model with specified properties, we can explain our results with explicit reasoning based on these properties.

\subsection{Scene Synthesis}
The general goal of indoor scene synthesis is to produce a feasible furniture layout of various object classes which address both functional and aesthetic criteria  \cite{zhang2019survey}. Early work of synthetic generation focused on hard-coding rules, guideline and grammars, resembling a procedural approach for this problem \cite{bukowski1995object, xu2002constraint, germer2009procedural}. The work of \cite{merrell2011interactive} is a successful example of hard-coding design guidelines as priors for the scene generation process. They extracted these guidelines through consulting manuals on furniture layout \cite{sharp2008art, ward1999use, talbott1999decorating} and interviewing professional designers who specialize in arranging furniture. A similar approach is also seen in Yu et al \cite{Yu2011} work, while \cite{yeh2012synthesizing} attempted synthesizing open world layouts with hard-coded factor graphs.

The work of  \cite{Fisher2012} can be seen as one of the early adapters of example-based scene synthesis. They synthesized scenes by training to build a probabilistic model based on Bayesian networks and Gaussian mixtures. Their problem, however, was one of generating an entire scene, and they utilized a more limited set of input example scenes. In the work of \cite{kermani2016learning}, a full 3D scene is synthesized iteratively by adding a single object at a time. This system learned some priors similar to ours, including pairwise and higher-order object relations. Compared  to this work, we incorporate additional priors, including objects’ relative position within the room bounds. The work of Liang et al. \cite{Liang2018, Liang2017} and Fu et al. \cite{fu2017adaptive} also took room functions into account. They argued that extracting topological priors should also be extended to room functions and their activities, which would impact the pair-wise relationships between objects. While object topologies differ in various room function, a major challenge in this approach is that not all spaces can be classified with a certain room function. For instance, in a small studio apartment, the living room might serve additional functions such as dining room and a study space. \cite{savva2017scenesuggest} also proposed a similar approach, involving a Gaussian mixture model and kernel density estimation. However, their system targeted an inverse problem of ours, namely, their problem received a selected object location as input and was asked to predict an object type. We find our problem to be more relevant to the needs of a content creator who knows what object they wish to place in scene, but does not have prior knowledge about the user’s surroundings.

Another data-driven approach to scene generation involves modeling human activities and interactions with the scene (\cite{fisher2015activity,ma2016action,fu2017adaptive,qi2018human}). Research following this approach generally seeks to model and adjust the entire scene according to human actions or presence. There have also been an number of interesting studies that take advantage of logical structures modeled for natural language processing (NLP) scenarios. Work of  \cite{chang2014learning}, \cite{chang2014interactive}, \cite{chang2017sceneseer} \cite{ma2018language} are examples of such approach. More specifically, \cite{ma2018language} bears a minor resemblance to our approach, in 1) training on object relations, and 2) the ability to augment an initial input scene, but unlike our work, it augments scenes by merging in subscenes retrieved from a database. In contrast, we seek to add in individual objects, which is more aligned with the needs of creators of augmented reality experiences.  A series of papers (including \cite{shao2012interactive,chen2014automatic,avetisyan2019scan2cad}) proposed generating a 3D scene representation by recreating the scene from RGB-D image input, using retrieved and aligned 3D models. This research, however, involves recreating an existing physical scene, and does not handle adding new objects. 

More recent work endeavors to improve learning-based methods, using deep convolutional priors \cite{wang2018deep}, scene-autoencoding \cite{li2019grains} and new representations of object semantics \cite{balint2019generalized}, to name just a few. \cite{zhang2020fast} addressed a related but distinct problem of synthesizing a scene by arranging and grouping an input set of objects. The work of Ritchie et al. \cite{ritchie2019fast} is another example of using deep generative models for scene synthesis. Their
method sampled each object attribute with a single inference
step to allow constrained scene synthesis. This work was extended in PlanIt \cite{wang2019planit}, where the authors proposed a combination of two separate convolutional networks to address constrained scene synthesis problems. They argue that object-level relationships facilitate high-level planning of how a room should
be laid out, while room-level relationships perform well at placing objects in precise spatial configurations. Our method differs from the discussed studies in 1) utilizing an explicit model rather than an implicit structure, 2) taking advantage of higher level relationships with the room itself in our proposed Scene Graph, and 3) generating a probability map which would guide the end user on potential locations for object augmentation.

\begin{figure}
  \includegraphics[width=1\columnwidth]{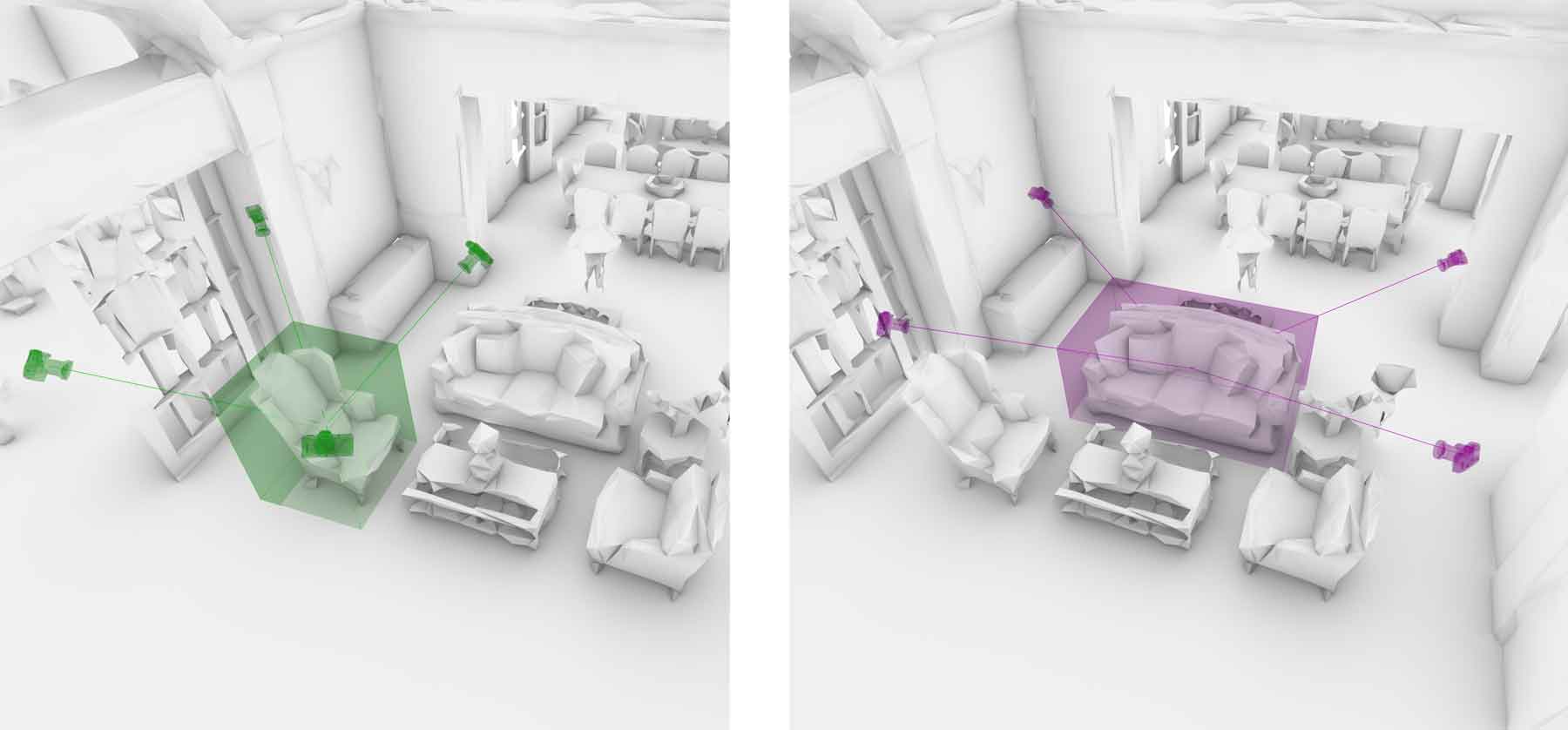}
  \caption[Camera Locations]{In our annotation tool, a camera is orbited around each object to facilitate labeling of object orientations. }~\label{fig:camLocation}
\end{figure}

\section{SceneGen Overview}
SceneGen is a framework to augment scenes with virtual objects using a generative model to maximize the likelihood of the relationships captured in a spatial Scene Graph. Specifically, if given a partially filled room, SceneGen, will augment it with one or multiple new virtual objects in a realistic manner using an explicit model trained on relationships between objects in the real world. The SceneGen workflow is shown in Figure \ref{fig:workflow}.

\begin{figure}
  \includegraphics[width=1\columnwidth]{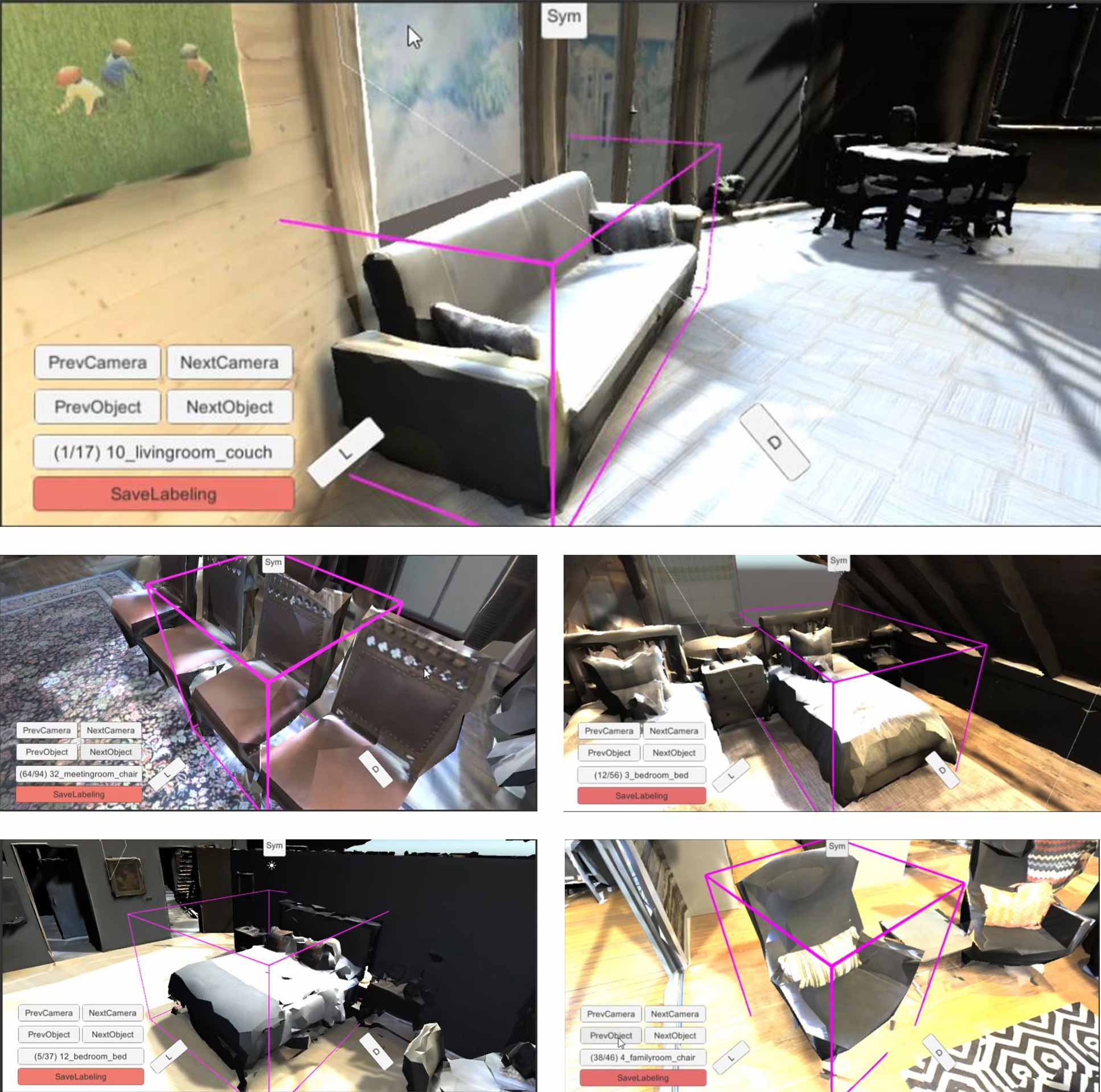}
  \caption[Orientation labeling GUI]{A labeler using our annotation tool can select which direction the object is facing or move to the next camera to get a better view. The selection is used to automatically standardize the axes of each object's bounding box.}~\label{fig:orilabel}
\end{figure}

\begin{figure*}
  \includegraphics[width=2\columnwidth]{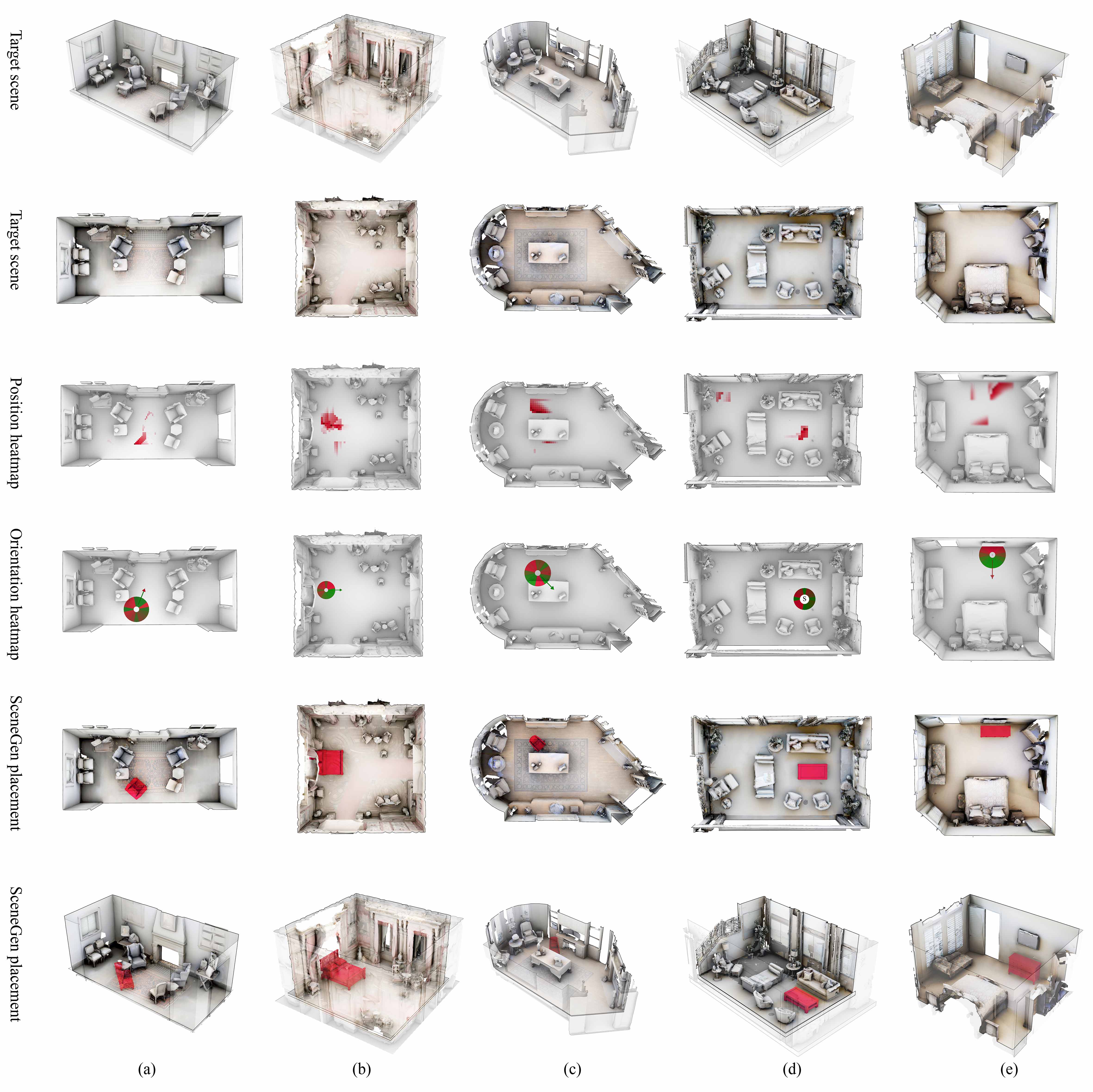}
  \caption[Single Object Augmentation]{Scene Gen places objects into scenes by extracting a Scene Graph from each room, sampling positions and orientations to create probability maps, and then places an object in the most probable pose. (a) A sofa is placed in a living room, (b) a bed is placed in a bedroom, (c) a chair is placed in an office, (d) A table is placed in a family room, (e) a storage is placed in a bedroom.}~\label{fig:together}
\end{figure*}

In this paper, we first introduce a novel Scene Graph that connects the objects and the room (both represented as nodes) using spatial relationships (represented as edges) in Section \ref{section:scene_rep}. For each object, these relationships are determined by positional and orientational features between itself and other objects, object groups, and the room.

In Section \ref{section:knowledge_model} we show how from a dataset of rooms, we can extract these Scene Graphs to construct a Knowledge Model that is used to train explicit models that approximate the probability density functions of position and orientation relationships for a given object using kernel density estimation. In order to augment a scene with a virtual object, SceneGen samples possible positions and orientations in a scene, building updated Scene Graphs for each sample. We estimate the probability of each sample and place an object at the most likely pose. SceneGen also shares a heat map of the likelihood of each sample to suggest alternate high probability placements. This can be repeated to augment multiple virtual objects.

Our implementation of SceneGen is built using data extracted from the Matterport3D dataset as our priors and is detailed in Section\ref{section:implementation}. This dataset is chosen as it contains real world rooms. As using object scans results in unoriented bounding boxes, we develop an application to facilitate the labeling of the facing direction of each object.

We assess the effectiveness of SceneGen in Sections \ref{section:experiments} and \ref{section:results} for eight categories of objects across several types of room including bedrooms, living rooms, hallways, and kitchens. In order to understand the effectiveness of each relationship on predicting where and how a new object should be placed, we run a series of ablation tests on each feature. We use k-fold cross validation to partition the Matterport3D dataset, building the Knowledge Model on a training set and assessing how well the model can replace removed objects from a validation set. Additionally, we carry out a user study to analyze how SceneGen compares with a random placement and the reference scene in placing new objects into virtual rooms based off of real scenes from the Matterport3D dataset and to evaluate the value of a heat map showing the probability of all samples. 

Finally, Section \ref{section:ar} details an Augmented Reality mobile application that we have developed employing SceneGen to add new virtual objects to a scene. This application locally computes the semantic segmentation and generates a Scene Graph before estimating sample probabilities on an external server, and then parses and visualizes the prediction results. This demonstrates how our framework can work with state-of-the-art AR/VR systems.

\section{Scene Representation}
\label{section:scene_rep}
\subsection{Graph Representation based on extracted features}

In this section, we introduce a novel spatial Scene Graph that converts a room and the objects included in it to a graphical representation using extracted spatial features. A Scene Graph $\mathcal{G}$ is defined by nodes representing objects, object groups, and the room, and by its edges representing the spatial relationships between the nodes. While various objects hold different individual functions (eg. a chair to sit, a table to dine, etc), their combinations and topological relationships tend to generate the main functional purpose of the space. In other words, spatial functions are created by the pair-wise topologies of objects and their relationship with the room. In our proposed Scene Graph representation, we intend to explicitly extract a wide variety of positional and orientational relationships that can be present between objects. We model descriptive topologies that are commonly utilized by architects and interior designers to generate spatial functionalities in a given space. Therefore, our Scene Graph representation can also be described as a function map, where objects (nodes) and their relationships (edges) correspond to a single or multiple spatial functionalities present in a scene. 
Figure \ref{fig:sceneGraph} illustrates two examples of our Scene Graph representation, where a subset of topological features are visualized in the graph.


\subsection{Definitions for Room and Objects}
\label{section:room_obj_defs}

In this paper, we consider a room or a scene in 3D space where its floor is on the flat $(x,y)$-plane and the $z$-axis is orthogonal to the $(x,y)$-plane. In this orientation,  we denote the room space in a floor-plan representation as $R$, namely, an orthographic projection of its 3D geometry plus a possible adjacency relationship that objects in $R$ may overlap on the $(x,y)$-plane but on top of one another along the $z$-axis. Specifically, the ``support'' relationship is defined in Section \ref{sec:support-definition}. This can also be viewed as a 2.5-D representation of the space. 

Further denote the $k$-th object (e.g., a bed or a table) in $R$ as $O_{k}$. The collection of all $n$ objects in $R$ is denoted as $\mathcal{O}= \{O_1, O_2,... O_n\}$.  $B(O_k)$ represents the bounding box of the object $O_k$. $\dot{O}_k$ represents the center of the object $O_k$. Every object $O_k$ has a label to classify its type. Related to the same $R$, we also have a set of groups $G$ = $\{g_1,..., g_m\}$, where each group $g_i$ contains all objects of the same type within $R$.

Furthermore, each $O_k$ has a primary axis $a_k$ and a secondary axis $b_k$. For Asymmetric objects, $a_k$ represents the orientation of the object. $a_k$ and $b_k$ are both unit vectors such that $b_k$ is a $\frac{\pi}{2}$ radian counter clockwise rotation of $a_k$.  We define $\theta_{a_k}$ and $\theta_{b_k}$ to be the angle in radians represented by $a_k$ and $b_k$ respectively.

 For each room $R$, we define $\mathcal{W} = \{W_1, W_2, ..., W_l\}$ where each $W_k$ is a wall of the $l$-sided room. In the floor plan representation, $W_k$ is represented by a 1D line segment. We also introduce a distance function $\delta(a, b)$ as the shortest distance between a and b objects. For example, $\delta(B(O_{k}),\dot{R})$ is the shortest distance between the bounding box of $O_k$ and the center of the room $R$.

\subsection{Positional Relationships}
We first introduce features for objects based on their spatial positions in a scene. We include both pairwise relationships between objects (eg. between a chair and a desk), object groups (eg. between a dining table and dining chairs), and relationships between an object and the room.


\subsubsection{Object to Room Relationships} \hfill\\
\textit{RoomPosition}: The room position feature of an object denotes whether an object is at the middle, edge, or corner of a room. This is based on how many walls an object is less than $\varrho$ distance from. 

\begin{equation}
    \label{eqn:room_position}
    \mbox{RoomPosition}\left(O_k, R\right) = \sum_{W_i \in \mathcal(W)} \mathbbm{1}(\delta (\dot{O}_k, W_i) < \varrho)  
\end{equation}
In other words, if $\mbox{RoomPosition}(O_k, R) \geq 2$, the object is near at least 2 walls of a room, and hence is near a \textit{corner} of the room; if $\mbox{RoomPosition}(O_k, R) = 1$, the object is near only one wall of the room and is at the \textit{edge} of the room; otherwise, the object is not near any wall and is in the \textit{middle} of the room.

\begin{figure*}
  \includegraphics[width=2\columnwidth]{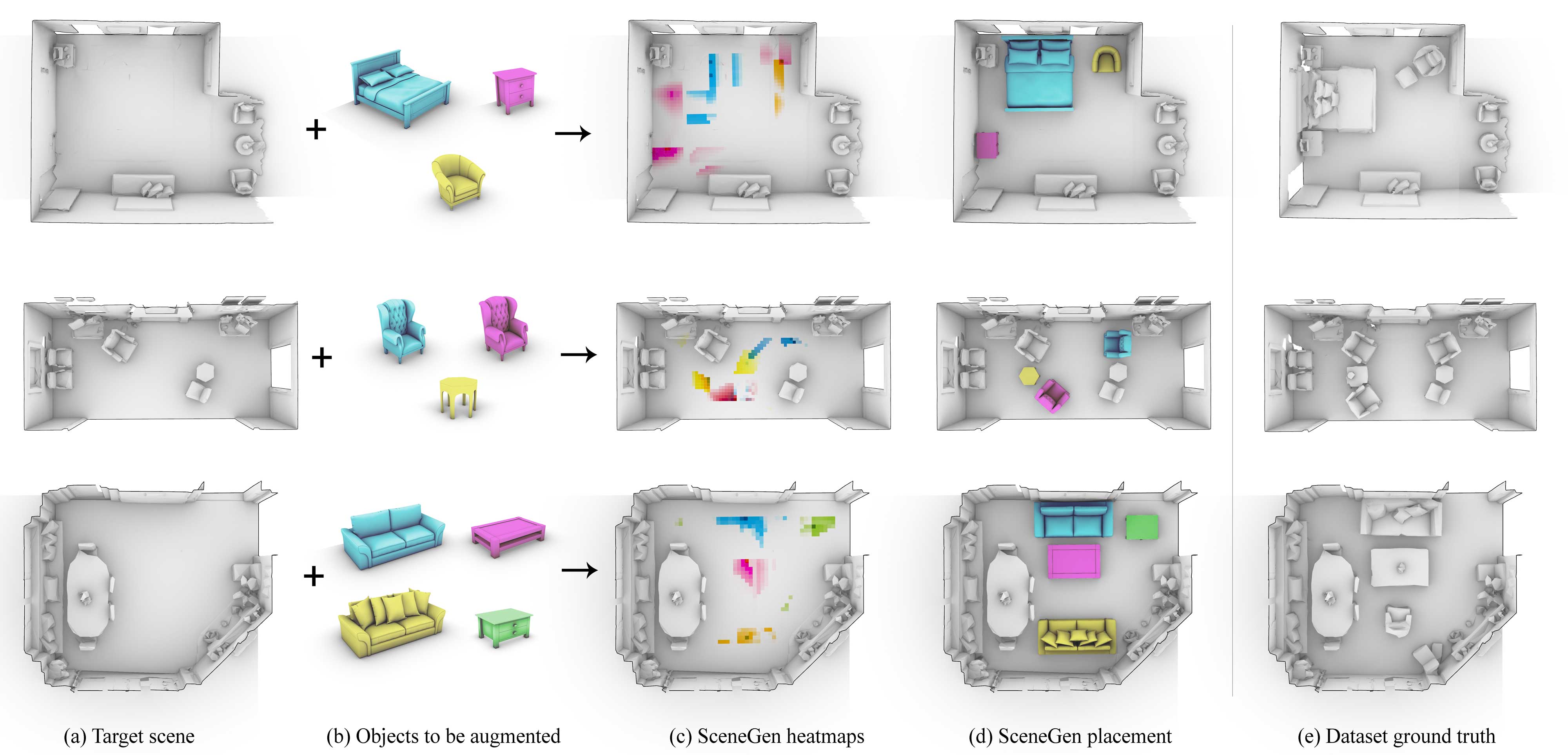}
  \caption{SceneGen can be used to iteratively add multiple virtual objects to a scene. For each object we sample poses and place it in the most likely position and orientation before placing the next object into a partially emptied room. (Top) A bed, storage and sofa are replaced in a bedroom, reorganizing the room in a viable alternative to the dataset ground truth; (Middle) Two sofas and a table are replaced to a living room in an arrangement similar to ground truth; (Bottom) A sofa, a table are replaced, and another sofa and then a table are added to a family room, demonstrating how a scene augmented with different objects compares to the ground truth.}~\label{fig:multiobject}
\end{figure*}
 
\subsubsection{Object to Object Group Relationships}\hfill\\
\textit{AverageDistance}: For each object, and each group  of objects we calculate the average distance between that object and all objects within that group. For cases where the object is a member of the group, we do not count the distance between the object in question and itself in the average. 
\begin{equation}
    \mbox{AverageDistance}(O_k, g_i) = 
    \sum\limits_{\substack{{O}_j \in g_i \\ j \neq k}\\} \delta(B(O_k), B(O_j)) / {\sum\limits_{\substack{O_j \in g_i \\ j \neq k}} 1}
\end{equation}
\textit{SurroundedBy}: For each object, and each group of objects, we compute how many objects in the group are within a distance $\varepsilon$ of the object.  For cases where the object is a member of the group, we do not count the object in question.
\begin{equation}
    \mbox{SurroundedBy}\left(O_k, g_i\right) =
    \sum\limits_{\substack{O_j \in g_i \\j \neq k}}  \mathbbm{1}( \delta\left(B(O_j), B(O_k)) < \varepsilon \right)
\end{equation}

\subsubsection{Object Support Relationships} \label{sec:support-definition}
\hfill\\
\textit{Support}: 
An object is considered to be supported by a group if is directly on top of an object from the group, or supports a group if it is directly underneath an object from the group.

\begin{equation}
    \mbox{Support}\left(O_k, g_i\right) = 
    \begin{cases} 
      1 & \exists O_j \in g_i \mbox{ where $O_k$ is on top of $O_j$} \\ 
      -1 & \exists O_j \in g_i \mbox{ where $O_k$ is under $O_j$}  \\
      0 & \text{otherwise}
   \end{cases}
\end{equation}

\subsection{Orientation Relationships}
\label{section:orientation_relationships}
We categorize the objects in our scenes into three main groups: 
\begin{enumerate}
    \item $G_{\mbox{sym}}$: Symmetric objects such as coffee tables and house plants that have no clear front-facing direction;
    \item $G_{\mbox{asym}}$: Asymmetric objects such as beds and chairs that can be oriented to face in a specific direction;
    \item $G_{\mbox{in}}$: Inside Facing objects such as paintings and storage that are always facing opposite to the wall of the room where they are situated.
\end{enumerate}
In this section we discuss features applicable to objects with a defined facing decisions, and not for symmetric objects.

\subsubsection{Object to Room Relationships}\hfill\\
We first define an indicator equation that is 1 if a ray extending from the center in the direction $d_k$ of an object intersects a wall $W_i$.
\begin{equation}
    f(\dot{O}_k, d_k, W_i) = \mathbbm{1}(\exists \gamma \geq 0 | \dot{O}_k + \gamma d_k \in W_i )
\end{equation}

\textit{TowardsCenter}: An object is considered to be facing towards the center of the room, if an ray extending from the center of the object intersects one of the furthest $\frac{l}{2}$ walls from the object.

\begin{equation}
    \label{eqn:towards_center}
    \begin{array}{l}
    c_1 = \argmax\limits_{W_i \in \mathcal(W)}{\delta (\dot{O}_k, W_i)} \\ 
    c_2 = \argmax\limits_{W_i \in \mathcal(W \setminus c_1)}{\delta (\dot{O}_k, W_i)} \\
    ...\\
    c_{\frac{l}{2}} = \argmax\limits_{W_i \in \mathcal(W \setminus c_1...c_{\frac{l}{2}-1})}{\delta (\dot{O}_k, W_i)}  \\
    \end{array}
\end{equation}
\begin{equation}
    \mbox{TowardsCenter}(O_k) = f(\dot{O}_k, a_k, c_1) \vee  ... \vee \mbox{f}(\dot{O}_k, a_k, c_{\frac{l}{2}-1})
\end{equation}

\begin{figure*}
  \includegraphics[width=2\columnwidth]{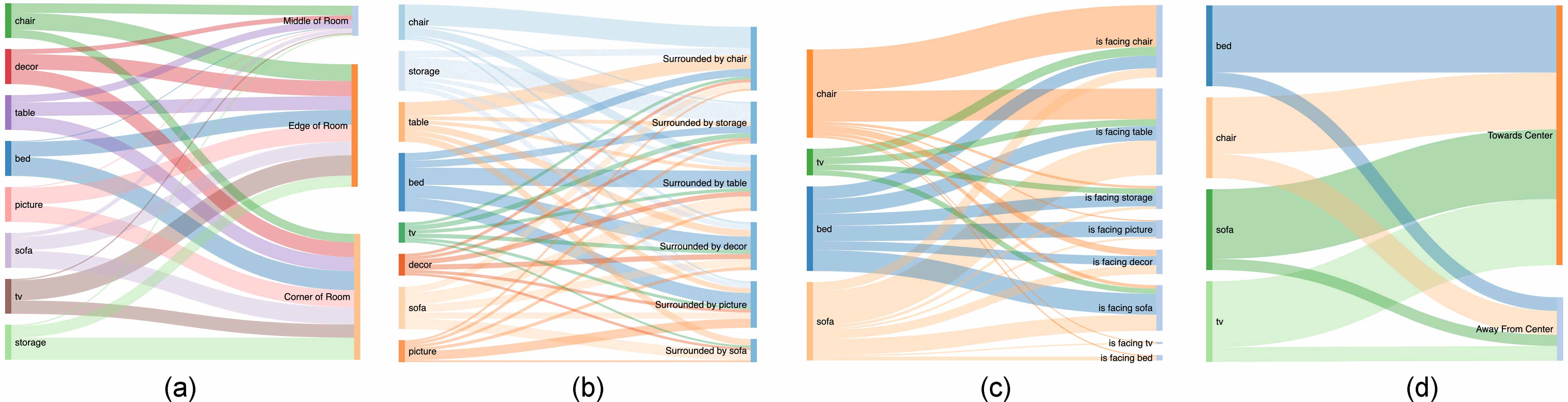}
  \caption[Relationships]{Visualization of the Knowledge Model built from Scene Graphs extracted from the Matterport3D Dataset shows for each  group of objects: (a) frequency of each Room Position, (b) frequency the object is surrounded by multiple objects from another group, (c) frequency the object is facing an object from another group, (d) frequency the object is facing towards the center of the room or not.}~\label{fig:relGraphs}
\end{figure*}

\textit{AwayFromWall}: An object is considered facing away from a wall if it is oriented away from and is normal to the closest wall to the object.
\begin{equation}
    \begin{array}{l}
    c_1 = \argmin\limits_{W_i \in \mathcal(W)}{\delta \left(B(O_k), W_i\right)}  \\ 
    \mbox{AwayFromWall}(O_k) = \mbox{f}(\dot{O}_k, -a_k, c_1) \wedge (a_k \perp c_i)
    \end{array}
\end{equation}

\textit{DirectionSimilarity}: An object has a similar direction as one or more objects within a constant $\varepsilon$ distance from the object if the other objects are facing in the same direction or in the opposite direction ($\pi$ radians apart) from the first object subject to some small angular error $\varphi$. 
\begin{equation}
\begin{array}{l}
    \mbox{Same}\left(O_k\right) = \sum\limits_{\substack{O_j \in \mathcal{O}, j \neq k \\ \delta \left(B(O_k), B(O_j)\right) \leq \varepsilon }} \mathbbm{1}(|\theta_{a_k} - \theta_{a_j}| \leq \varphi) \\
    \mbox{Opp}\left(O_k\right) = \sum\limits_{\substack{O_j \in \mathcal{O}, j \neq k\\ \delta\left(B(O_k), B(O_j)\right) \leq \varepsilon}} \mathbbm{1}(|\pi -|\theta_{a_k} - \theta_{a_j}|| \leq \varphi)  \\
    \mbox{DirectionSimilarity}(O_k) = [\mbox{Same}(O_k), \mbox{Opp}(O_k)] \in \mathbb{R}^2
\end{array}
\end{equation}

\subsubsection{Object to Object Group Relationships}\hfill\\
We first define an indicator function that is 1 if a ray extending from the center of the object in direction $d_k$ intersects the bounding box of a second object.
\begin{equation}
    h(\dot{O}_k, d_k, B(O_j)) = \mathbbm{1}(\exists \gamma \geq 0 | \dot{O}_k + \gamma d_k \in B(O_j))
\end{equation}
\textit{Facing}: Between an object and a group of objects we count how many objects of the group are within a distance $\varepsilon$ of the object and are in the direction of the primary axis of the first object.
\begin{equation}
   \mbox{Facing}(O_k, g_i) = \sum_{\substack{O_j \in g_i, j \neq k\\ \delta(B(O_k), B(O_j)) \leq \varepsilon}}
   h(\dot{O}_k, a_k, B(O_j))
\end{equation}

\textit{NextTo}: Between an object and a group of object we count how many objects of the group are within a distance $\varepsilon$ of the object and are in the direction of the positive or negative secondary axis of the first object.
\begin{equation}
   \mbox{NextTo}\left(O_k, g_i\right) = \sum_{\substack{O_j \in g_i, j \neq k \\ \delta \left(B(O_k), B(O_j)\right) \leq \varepsilon}}
   h\left(\dot{O}_k, \pm b_k, B(O_j)\right)
\end{equation}

\section{Knowledge Model}
\label{section:knowledge_model}
\subsection{Feature Vectors for Position and Orientation}
To evaluate the plausibility of a new arrangement, we compare its corresponding Scene Graph with a population of viable Scene Graphs priors. By extracting Scene Graphs from a corpus of rooms, we construct a Knowledge Model which serves as our spatial priors for the position and orientation relationships of each object group. For each object instance, we assemble a data vector for positional features from $\mathcal{G}$. For Asymmetric objects, we similarly create a data vector for orientational features. First we define the following that represent an object's relationships with all groups, G = \{$g_1,..., g_m$\}.  
\begin{equation}
    \label{eqn:object_group_relationships}
    \begin{array}{l}
    \mbox{AD}\left(O_k\right) = [\mbox{AverageDistance}(O_k, g_i) | i = 1,\cdots, m] \in \mathbb{R}^m  \\
    \mbox{S}\left(O_k\right) = [\mbox{SurroundedBy}(O_k, g_i) | i = 1,\cdots, m]\in \mathbb{R}^m  \\
    \mbox{F}\left(O_k\right) = [\mbox{Facing}(O_k, g_i) | i = 1,\cdots, m]\in \mathbb{R}^m  \\
    \mbox{NT}\left(O_k\right) = [\mbox{NextTo}(O_k, g_i)| i = 1,\cdots, m]\in \mathbb{R}^m  \\
    {\mbox{SP}}\left(O_k\right) = [\mbox{Support}(O_k, g_i) | i = 1,\cdots, m]\in \mathbb{R}^m  \\

    \end{array}
\end{equation}
This allows us to construct data arrays,  $d_p\left(O_k\right)$ and $d_o\left(O_k\right)$, containing features that relate to the position and orientation of an objects respectively. \mbox{RoomPosition} is also included in the data array for orientational features, $d_o$, since the other features of $d_o$ are strongly correlated with an object's position in the room. This is abbreviated as \mbox{RP}. We also use the abbreviate \mbox{TowardsCenter} to \mbox{TC} and \mbox{DirectionSimilarity} to \mbox{DS}. For succinctness, when using these abbreviations for our features, the parameter $O_k$ is dropped from our notation.

\begin{equation}
    \label{eqn:object_relationships}
    \begin{array}{l}
        d_p\left(O_k\right) = [\mbox{RP}\in\mathbb{R}, \mbox{AD}\in\mathbb{R}^m, \mbox{SP}\in\mathbb{R}^m, \mbox{S}\in\mathbb{R}^m] \in \mathbb{R}^{3m +1} \\
        d_o\left(O_k\right) = [\mbox{RP}\in\mathbb{R}, \mbox{TC}\in\mathbb{R}, \mbox{DS}\in\mathbb{R}^2,\mbox{F}\in\mathbb{R}^m, \mbox{NT}\in\mathbb{R}^m] \in \mathbb{R}^{2m +4}
    \end{array}
\end{equation}

Finally, given one feature vector per object for position and orientation, respectively, we can collect more samples from a database, which we will discuss in Section \ref{section:implementation}, to form our Knowledge Model. The model collects feature vectors separately with respect to different object types in multiple room spaces. To do so, we introduce $g_{i,j}$ to collect all of the $i$-th type objects in room $R_j, j=1, \cdots, r$. Without loss of generality, we assume that the $i$-th object type is the same across all rooms. Therefore, we can collect all the objects of the same $i$-th type from a database as
$$
g_{i,*} = \bigcup_{j=1}^{r} g_{i,j}. 
$$
Then $D_p(g_{i,*})$ and $D_o(g_{i,*})$ represent the collections of all feature vectors in \eqref{eqn:object_relationships} from objects in $g_{i,*}$.

\begin{equation}
    \label{eqn:dataset_relationships}
    \begin{array}{l}
        \mathcal{D}_p(g_{i,*}) = \{d_p\left(O_k\right)| \forall O_k \in g_{i,*}\}  \\
        \mathcal{D}_o(g_{i,*}) = \{d_o\left(O_k\right)| \forall O_k \in g_{i,*}\} 
    \end{array}
\end{equation}

\subsection{Scene Augmentation}
%
%
%

Given the feature samples for the same type of object in \eqref{eqn:dataset_relationships}, now we can estimate their likelihood distribution. In particular, given an object placement $O$ of the $i$-th type, we seek to estimate the likelihood function for its position features:
\begin{equation} \label{eqn:P-d-p}
    P(d_p(O)|\mathcal{D}_p (g_{i,*} )).    
\end{equation}
If $O$ is asymmetric, we also seek to estimate the likelihood function for its orientation features:
\begin{equation}\label{eqn:P-d-o}
    P(d_o(O)| \mathcal{D}_o (g_{i,*})).
\end{equation}
However, if $O$ is an Inside Facing object, then with certainty its orientation will be determined by that of its adjacent wall. Additionally, if $O$ is a Symmetric object, it has no clear orientation. Therefore, for these categories of objects, estimation of their orientation likelihood is not needed. In this section, we discuss how to estimate \eqref{eqn:P-d-p} and \eqref{eqn:P-d-o}

We can approximate the shape of these distributions using multivariate kernel density estimation (KDE). Kernel density estimation is a non-parametric way to create a smooth function approximating the true distribution by summing kernel functions, $K$, placed at each observation $X_i...X_n$ \cite{DensityEstimation}. 
\begin{equation}
    \hat{f}_h(x) = \frac{1}{nh} \sum_{i = 1}^{n} K\left(\frac{x-X_i}{h}\right)
\end{equation}

This allows us to estimate the probability distribution function (PDF) of the position and orientation relationships from the spatial priors in our Knowledge Model, $\mathcal{D}_p(g_{i,*}),\mathcal{D}_o(g_{i,*})$ for each group $g_i$. 

\subsection{SceneGen Algorithm}

Algorithm \ref{alg:SceneGen} describes the SceneGen algorithm. Given a room model $R$ and a set of existing objects $\mathcal{O}= \{O_1, O_2,... O_n\}$, the algorithm evaluates the position and orientation likelihood of augmenting a new object $O'$ and recommends its most likely poses.
\begin{algorithm}[h]
\caption{SceneGen Algorithm} \label{alg:SceneGen}
Given a training database, calculate $\mathcal{D}_p(g_{i,*})$ and $\mathcal{D}_o(g_{i,*}) $ as prior.

For a given room $R$, construct the Scene Graph $\mathcal{G}$ of its objects $\mathcal{O}$.

\While{Sample the position of $O'$ of type $i$ in $R$} {

    Calculate $ P(d_p(O')|\mathcal{D}_p (g_{i,*} ))$.
     
    \While{Sample the orientation of $O'$ $\in [0, 2\pi)$ } {
        
        Calculate $ P(d_o(O')| \mathcal{D}_o (g_{i,*}))$
    }
}

Generate a heat map displaying the likelihood distributions.

Make recommendation to place $O'$ at the highest probability pose.

\end{algorithm}

Figure \ref{fig:graphoptions} shows how potential scene graphs are created for sampled placements. For scenes where multiple objects need to be added, we repeat Algorithm \ref{alg:SceneGen} for each additional object.

\section{Implementation}
\label{section:implementation}
In this section, we discuss the implementation detail of SceneGen framework based on the relationship data learned from the Matterport3D dataset. 

\subsection{Dataset}
Matterport3D {\cite{Chang2018}} is a large-scale RGB-D dataset containing 90 building-scale scenes. The dataset consists of various building types with diverse architecture styles, including numerous spatial functionalities and furniture layouts. Annotations of building elements and furniture have been provided with surface reconstruction as well as 2D and 3D semantic segmentation.

\subsubsection{Pose Standardization}
In order to use the Matterport3D dataset as priors for SceneGen, we must make a few modifications to standardize object orientations using an annotation tool we have also developed. In particular, different from Section \ref{section:room_obj_defs}, our annotation tool interacting with the dataset is fully in 3D environment (i.e., through Unity 3D). After the annotation, the relationship data then are consolidated back to the 2.5-D representation conforming to the computation of the SceneGen models.

For each object $O_k$, the Matterport3D dataset supplies labeled oriented 3D bounding boxes $B(O)$ aligned to the $(x,y)$-plane. This is defined by a center position $\dot{O}$, primary axis $a$, secondary axis $b$, an implicit tertiary axis $c$, and $r \in \mathbb{R}^3 $ denotes the bounding box size of $O$ divided in half. However, the Matterport3D dataset does not provide information about which labeled direction the object is facing or aligns with the $z$-axis. Hence, it will rely on our labeling tool to resolve the ambiguities.

To provide a consistent definition, we describe a scheme to label these axes such that the primary axis, $a$ points in the direction the object is facing, $a^*$. Since we know that only one of these three axes has a $z$ component, we shall store this in the third axis $c$ and define $b$ to be orthogonal to $a$ on the $x,y$ plane. The box size $r$ will also be updated to correspond to the correct axes. By constraining these axes to be right handed, for a given $a^*$ we have: 
\begin{equation}
  \label{eqn:relabel_scheme}
  \begin{array}{l}
    c^* \doteq [0,0,1], \quad b^* \doteq c^* \times a^*.
    \end{array}
\end{equation}

In order to correctly relabel each object, we have developed an application to facilitate the identification of the correct primary axis for all Asymmetric objects and supplemented this to the updated data set.

For each object, we view the house model mesh at different camera positions around the bounding box in order to determine the primary axis of the object as displayed in Figure \ref{fig:camLocation}. Our annotation tool shown in Figure \ref{fig:orilabel} allows a labeler to select from two possible directions at each camera position or can move the camera clockwise or counter clockwise to get a better view.  Once a selection is made, the orienting axis, $a^*$ can be determined by knowing which camera we are looking at and the direction selected. We use \eqref{eqn:relabel_scheme} to standardize the axes. Using our annotation tool, the orientations of all objects in a typical house scan can be labeled in about 5 minutes.

\begin{figure}
  \includegraphics[width=0.80\columnwidth]{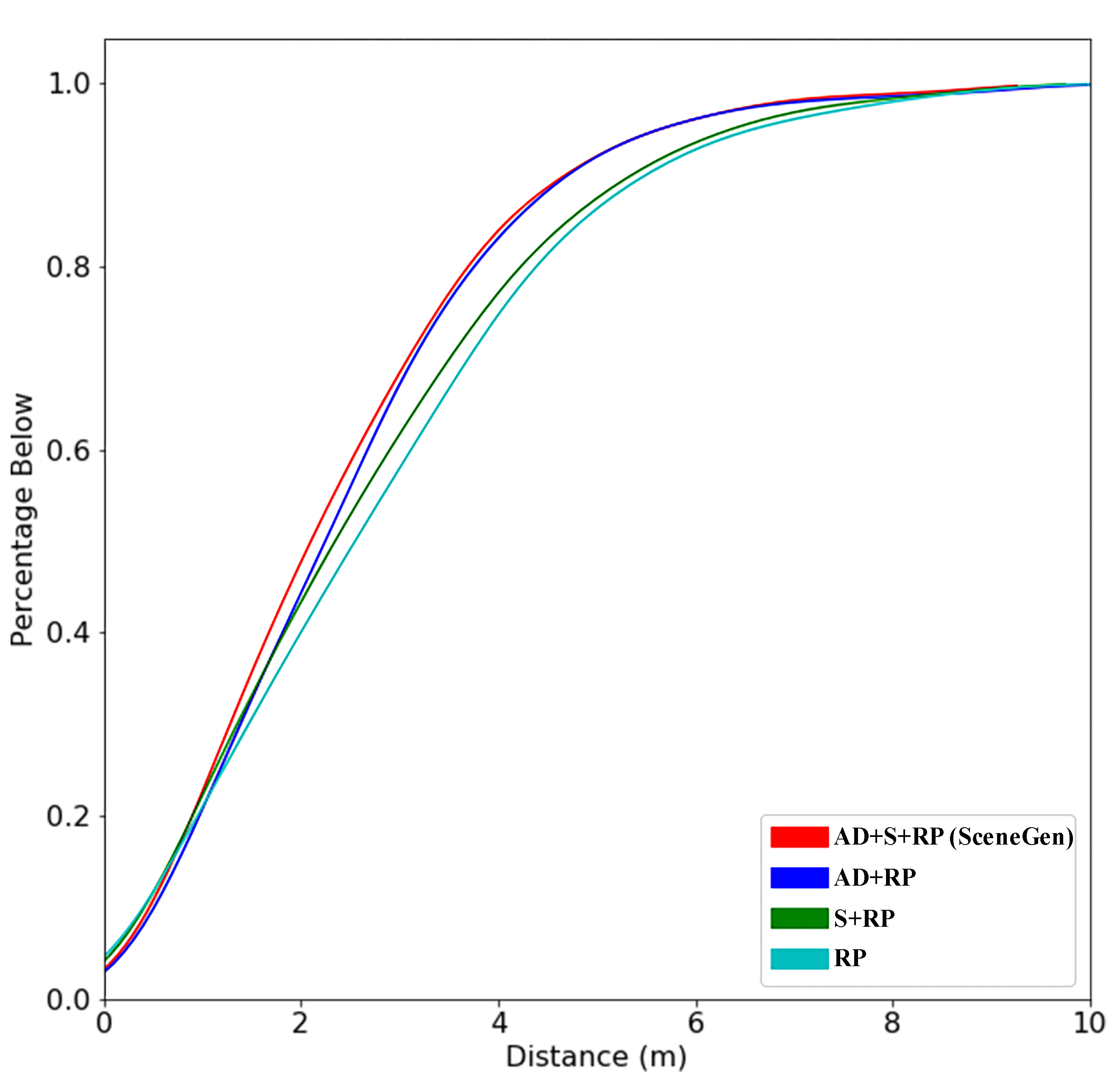}
  \caption{Distance between the ground truth object's position and where SceneGen and other ablated versions of our system predicts the object should be re-positioned is shown in a cumulative density plot. }~\label{fig:abl_pos1}
\end{figure}

\begin{figure}
  \includegraphics[width=0.80\columnwidth]{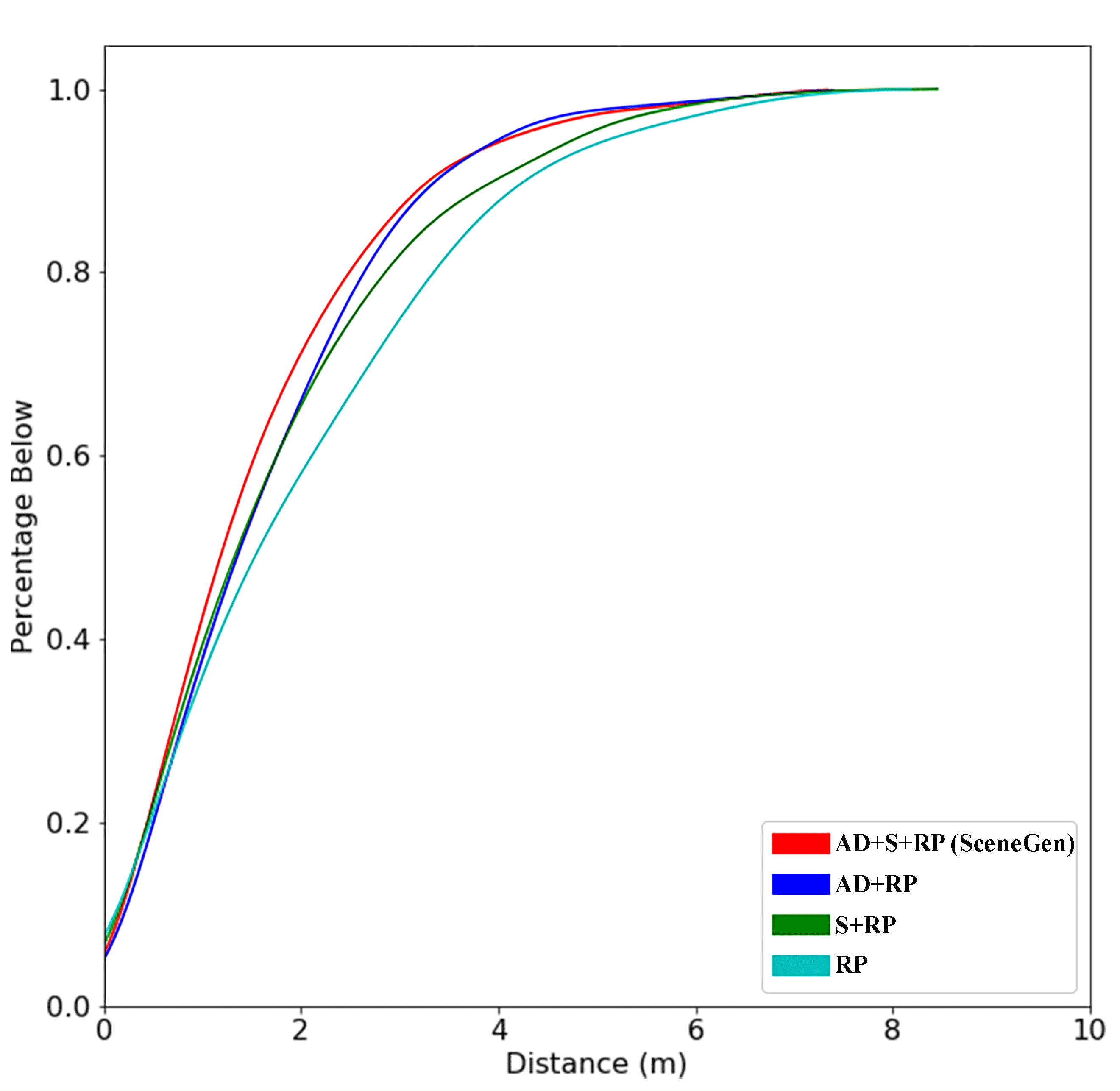}
  \caption{Distance between the ground truth object's position and the nearest of the 5 highest probability positions predicted by SceneGen and other ablated versions of our system is shown in a cumulative density plot.}~\label{fig:abl_pos2}
\end{figure}



\begingroup
\setlength{\tabcolsep}{3pt} 
\renewcommand{\arraystretch}{1} 
\begin{table*}

  \caption{Distance between ground truth and predicted positions for different models, with smallest distances for each object type in bold (ablation study). Topology features are abbreviated as follows: AverageDistance as AD, SurroundedBy as S, and RoomPosition as RP.}\label{tab:table_ablation_pos}
  \centering
  \begin{tabular}{|c| c c | c c| c c | c c | c c | c c | c c | c c | c c|}
\hline
    {\textit{System}}
    & \multicolumn{2}{c}{ \textit{Bed}}
      & \multicolumn{2}{c}{ \textit{Chair}}
    & \multicolumn{2}{c}{ \textit{Storage}}
     & \multicolumn{2}{c}{\textit{Decor}}
      & \multicolumn{2}{c}{ \textit{Picture}}
       & \multicolumn{2}{c}{ \textit{Table}}
        & \multicolumn{2}{c}{ \textit{Sofa}}
         & \multicolumn{2}{c}{\ \textit{TV}}
          & \multicolumn{2}{c}{ \textit{Overall}}\\
          \hline 
          \hline
    
    & 
 \footnotesize{Top 1} & \footnotesize{Top 5}&
 \footnotesize{Top 1} & \footnotesize{Top 5}&
 \footnotesize{Top 1} & \footnotesize{Top 5}&
 \footnotesize{Top 1} & \footnotesize{Top 5}&
 \footnotesize{Top 1} & \footnotesize{Top 5}&
 \footnotesize{Top 1} & \footnotesize{Top 5}&
 \footnotesize{Top 1} & \footnotesize{Top 5}&
 \footnotesize{Top 1} & \footnotesize{Top 5}&
 \footnotesize{Top 1} & \footnotesize{Top 5}\\
     \hline

         AD+S+RP (SceneGen) &
    1.58& \textbf{0.87}& \textbf{2.26}& \textbf{1.35}& \textbf{2.27}& \textbf{1.45}& \textbf{2.71}& \textbf{1.71}& \textbf{2.80}& 1.99& \textbf{2.15}& 1.47& 2.56& 1.58 &2.49& \textbf{1.52}& \textbf{2.40}& \textbf{1.54} \\

         AD + RP &
    \textbf{1.40}& 0.95& 2.40& 1.47& 2.55& 1.67& 2.79& 1.96& 2.95& 2.03& 2.26& \textbf{1.46}& 2.58& \textbf{1.58}& \textbf{2.39}& 1.731& 2.49& 1.65 \\
        S + RP &
    1.85& 1.32& 2.46& 1.56& 2.46& 1.64& 3.38& 2.14& 2.82& \textbf{1.92}& 2.67& 1.72& \textbf{2.53}& 1.64& 2.51& 1.55& 2.67& 1.73 \\
        RP &
    1.99& 1.31& 2.95& 2.31& 2.75& 1.53& 3.12& 2.56&2.95& 2.21& 2.70& 1.57& 2.55& 1.72& 2.95& 2.32& 2.80& 1.96 \\
     \hline
  \end{tabular}
\end{table*}
\endgroup

\subsubsection{Category reduction}

For this study, we have reduced the categories of object types considered for building our model and placing new objects. Though the Matterport3D dataset includes many different types of furniture, organized with room labels to describe furniture function (e.g. "dining chair" v.s. "office chair"), we found that the dataset has a limited amount of instances for many object categories. Because we build statistical models for each object category, we require an adequate representation of each category. Thus, we reduce the categories to a better-represented subset for the purposes of this study.

We group the objects into 9 broader categories: $G =$ \{Bed, Chair,  Decor, Picture, Sofa, Storage, Table, TV, Other \}. Each of these categories has a specific type of orientation, as described in Section \ref{section:orientation_relationships}. Of these categories, Asymmetric objects are $G_{\mbox{asym}} = \{\mbox{Bed, Chair, Sofa, TV}\}$, Symmetric objects are $G_{\mbox{sym}} = \{\mbox{Decor, Table}\}$, and Inside Facing objects are $G_{\mbox{in}} = \{ \mbox{Picture, Storage}\}$. 

For room types, we consider the set $\{$ library,  living room, meeting room, TV room, bedroom, rec room, office, dining room, family room, kitchen, lounge$\}$ to avoid overly specialized rooms such as balconies, garages and stairs. We also manually eliminate unusually small or large rooms with outlier areas and rooms where scans and bounding boxes are incorrect. 

After the data reduction, we consider a total of 1,326 rooms and 7,017 objects in our training and validation sets. The object and room categories used can be expanded if sufficient data is available.

\subsection{Knowledge Model}
We use the processed dataset as prior to train the SceneGen Knowledge Model. The procedure first estimates each object $O_k$ according to \eqref{eqn:object_relationships}, and subsequently constructs $\mathcal{D}_p(g_{i,*})$ and $\mathcal{D}_o(g_{i,*})$ in \eqref{eqn:dataset_relationships} for categories in $G$ and $G_{\mbox{asym}}$ respectively. We do not construct models for the `Other' category as objects contained in this category are sparse and unrelated from each other. Given our priors, we estimate the likelihood functions $P(d_p(O)|\mathcal{D}_p(g_{i,*}))$ and $P(d_o(O)|\mathcal{D}_p(g_{i,*}))$ from \eqref{eqn:P-d-p} and \eqref{eqn:P-d-o} using Kernel Density Estimation. 

We utilize a KDE library developed by \cite{seabold2010statsmodels} with a normal reference rule of thumb bandwidth with ordered, discrete variable types. We make an exception for AverageDistance, which is continuous. When there are no objects of a certain group, $g_i$ in a room, the value of $\mbox{AverageDistance}(O_k, g_i)$ is set to a large constant (1000), and we use a manually tuned bandwidth (0.1) to reduce the impact of this on the rest of the distribution. 


Furthermore, we found for this particular dataset, a subset of features, Facing, TowardsCenter and RoomPosition, are most impactful in predicting orientation as detailed in Section \ref{section:orientation_ablation_results}. Therefore, while we model all of the orientational features, we only use the Facing, TowardsCenter and RoomPosition features for our implementation of SceneGen and in the User Studies. Finally, due to overlapping bounding boxes in the dataset, calculating object support relationships (SP) precisely is not possible. Thus in our implementation, we allow the certain natural overlaps defined heuristically instead of using these features. A visualization of our priors from the Matterport3D dataset can be seen in Figure \ref{fig:relGraphs}.


We use Algorithm \ref{alg:SceneGen} to augment a room $R$ with an object of type $i$ and generate a probability heat map. This can be repeated in order to add multiple objects.  To speed up computation in this implementation, we first sample positions, and then sample orientations at the most probable position, instead of sampling orientations at every possible position. 

Figure \ref{fig:together} shows how our implementation of SceneGen adds a new object to a scene and examples of scenes are augmented with multiple objects iteratively is shown in Figure \ref{fig:multiobject}.

\paragraph{Computation Time}
We train and evaluate our model using a machine with an 4-core Intel i7-4770HQ CPU and 16GB of RAM. In training, creating our Knowledge Model and estimating distributions for 8 categories of objects takes approximately 12 seconds. In testing, it takes $\approx$2 seconds to extract a scene graph and generate a heat map indicating the probabilities of 250 sampled poses.

\begin{figure}
  \includegraphics[width=0.80\columnwidth]{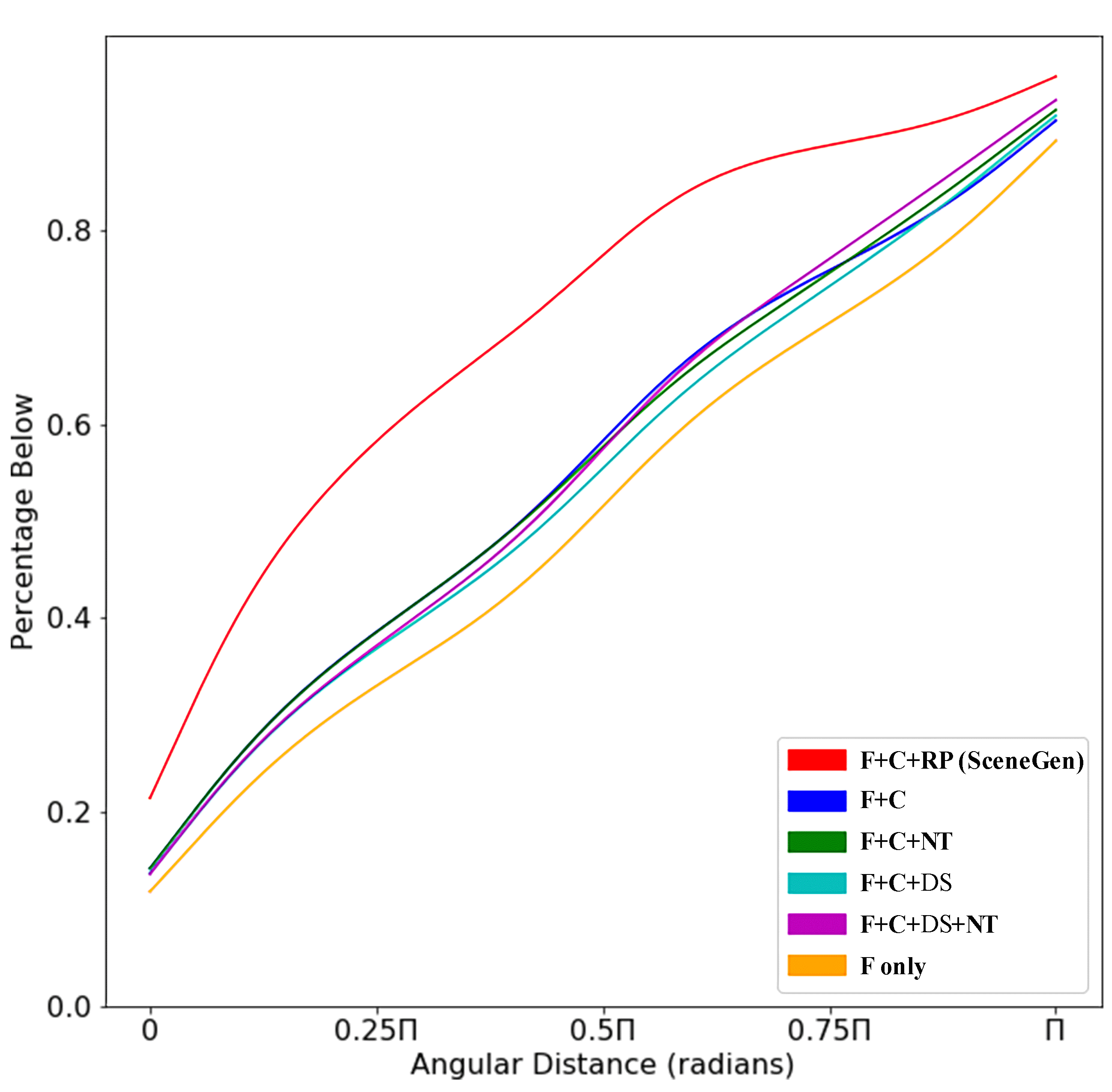}
  \caption{Cumulative density plot indicates angular distance between ground truth orientation and our system's predicted orientation for SceneGen and other subsets of orientation features. The range is $[0, \pi)$.}~\label{fig:abl_or}
\end{figure}

\begingroup
\setlength{\tabcolsep}{3pt} 
\renewcommand{\arraystretch}{1} 
\begin{table}
  \caption{Angular Distance between ground truth and predicted orientations for different model architectures (ablation study). Topology features are abbreviated as follows: Facing as F, TowardsCenter as C, RoomPosition as (RP), NextTo as NT, DirectionSimilarity as DS. }\label{tab:table_ablation_orien}
  \centering

\begin{tabular}{|l|l|l|l|l|l|}
\hline

\textit{System} & \textit{ Bed}   & \textit{ Chair} & \textit{ Sofa}  & \textit{ TV}    & \textit{ Overall} \\
 \hline
 \midrule

 { F+C+RP (SceneGen)} & 
\textbf{0.65} & \textbf{ 0.98} & \textbf{ 0.67} & {0.66} & \textbf{ 0.85}  \\
{ F only } & { 1.13  } & { 1.66} & { 1.51} & { 0.91}  & { 1.54}   \\
{ F+C} & { 1.13} & { 1.55} & { 1.18} & { 0.49} & { 1.35}   \\
{ F+C+NT} & { 1.18} & { 1.53} & { 1.23} & \textbf{0.46} & { 1.35}   \\
{ F+C+DS} & { 1.54} & { 1.55} & { 1.21} & { 0.59} & { 1.39}   \\
{ F+C+DS+NT} & { 1.22} & { 1.50} & { 1.23} & { 0.63} & { 1.35}  \\
\hline

\end{tabular}
\end{table}
\endgroup

\section{Experiments}
\label{section:experiments}

\subsection{Ablation Studies}

To evaluate our prediction system, we run ablation studies, examining how the presence or absence of particular features affects our object position and orientation prediction results. We use a K=4-fold cross validation method on our ablation studies, with 100 rooms in each validation set and the remaining rooms in our training set. 

\begin{figure*}
  \includegraphics[width=2.0\columnwidth]{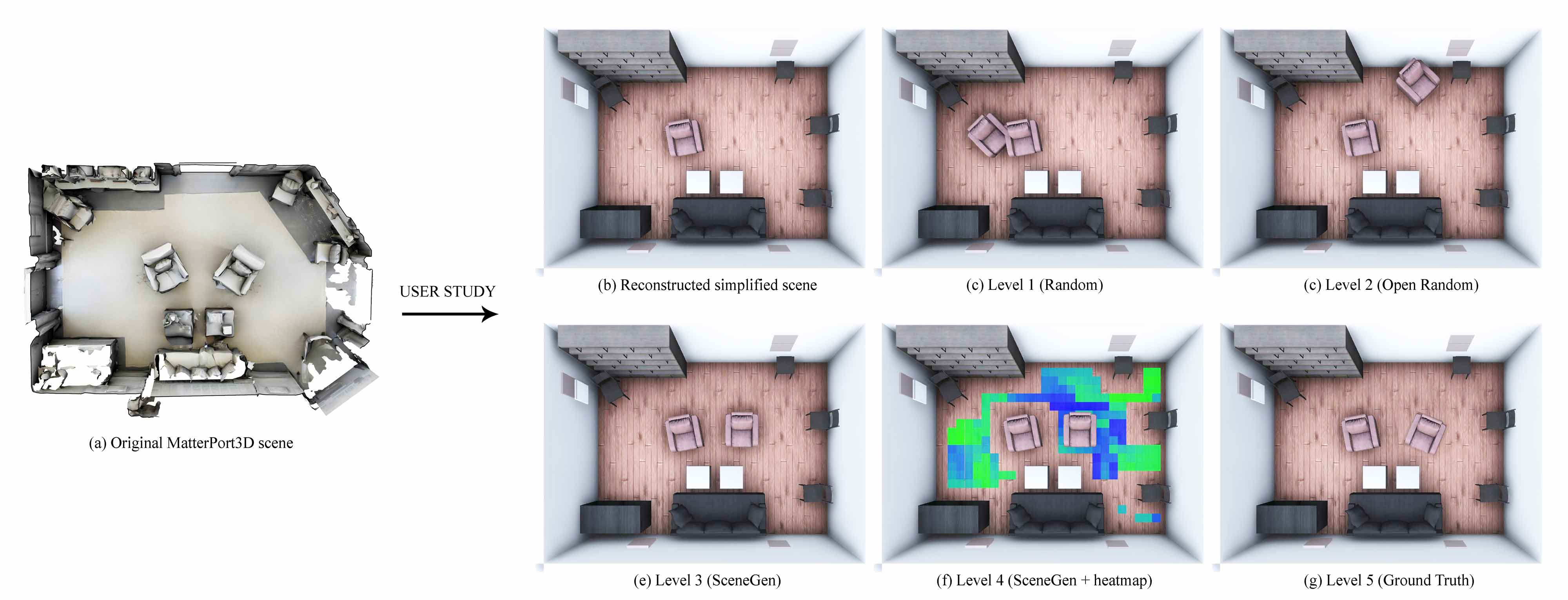}
  \caption{Users are shown scenes that are simplified models based on original Matterport3D rooms. An object is replaced in rooms using one of 5 levels of the systems. Level I places the object randomly in the room. Level II places the object randomly in an open space. Levels III and IV use SceneGen to predict the most likely placement and orientation, and Level IV also shows a heat map visualizing the probabilities of each sampled position. In Level V, the user sees the ground truth scene. When viewing the 3D model during experiment, the user have multiple camera angles available and is able to pan, zoom and orbit around the 3D room to evaluate the placement.}~\label{fig:userStudyLev}
\end{figure*}

\subsubsection{Position Features Evaluation}

The full position prediction model, SceneGen, trains three features: AverageDistance (AD), SurroundedBy (S), RoomPosition (RP) or AD+S+RP in short.  We create three reduced versions of our system: AD+RP, using only AverageDistance and RoomPosition features; S+RP, using only Surrounding and RoomPosition features; and RP, solely using the RoomPosition feature. 

We evaluate each system using the K-fold method described above. In this study, we remove each object in the validation set, one at a time, and use our model to predict where the removed object should be positioned. The orientation of the replaced object will be the same as the original. We compute the distance between the original object location and our system's prediction.  
 
 However, as inhabitants of actual rooms, we are aware that there is often more than one plausible placement of an object, though some may be more optimal than others. Thus, we raise the question of whether there is more than one ground truth or correct answer for our object placement problem. Hence, in addition to validating our model's features, our first ablation study validates them in relation to the simple approach of taking the single highest-scored location from our system. Meanwhile, our second ablation study uses the top 5 highest-scored locations, opening up examination to multiple potential "right answers". 

\subsubsection{Orientation Features Evaluation}

We run a similar experiment to evaluate our orientation prediction models for Asymmetric objects. Our Scene Graphs capture 5 relationships based on the orientation of the objects: Facing (F), TowardsCenter (C), NextTo (NT),  DirectionSimilarity (DS), and RoomPosition (RP). We assess models based on several combinations of these relationships.

We evaluate each of these models using the same K-fold approach, removing the orientation information of each object in the validation set, and then using our system to predict the best orientation, keeping the object's position constant. We measure the angular distance between our system's predictions and the original object's orientation.

\subsection{User Evaluation}
We conduct user studies with a designed 3D application based on our prediction system to evaluate the plausibility of our predicted positions and the usefulness of our heat map system. We recruited 40 participants, of which 8 were trained architects. To ensure unbiased results, the participants were randomly divided into 4 groups. Each group of users were shown 5 scenes from each of the 5 levels for a total of 25 scenes. The order these scenes were presented in was randomized for each user and they were not told which level a scene was at. 

We reconstructed 34 3D scenes from our dataset test split, where each scene had one object randomly removed. In this reconstruction process, we performed some simplification and regularized the furniture designs using prefabricated libraries, so that users would evaluate the layout of the room, rather than the design of the object itself, while matching the placement and size of each object. An example of this scene reconstruction and simplification can be seen in Figure \ref{fig:userStudyLev}(a-b).

The five defined levels test different object placement methods as shown in Figure \ref{fig:userStudyLev}(c-g) to replace the removed object. Levels I and II are both random placements, generated at run time for each user. The Level I system initially places the object in a random position and orientation in the scene. The Level II system places the object in an open random position and orientation, where the placement does not overlap with the room walls or other objects. Levels III and IV use SceneGen predictions. The Level III system places the object in the position and orientation predicted by SceneGen. Level IV also places the object in the predicted position and orientation, but also overlays a probability map. The Level V system places the object at the position it appears in the Matterport3D dataset, i.e., the ground truth.

We recorded the users' Likert rating of the plausibility of the initial object placement on a scale of 1 to 5 (1 = implausible/random,3 = somewhat plausible, 5 = very plausible). We also recorded whether the user chose to adjust the initial placement, the Euclidean distance between the initial placement and the final user-chosen placement, the orientation change between the initial orientation and the final user-chosen orientation. We expect higher initial Likert ratings and smaller adjustments to position and orientation for levels initialized by our system than for levels initialized to random positions.

Each participant used an executable application on a desktop computer. The goal of the study was explained to the user and they were shown a demonstration of how to use the interface. For each scene, the user was shown a 3D room and an object that was removed. After inspecting the initial scene and clicking "place object", the object was placed in the scene using the method corresponding to the level of the scene. In Level IV Scenes, the probability heat map was also visualized. The user was shown multiple camera angles and was able to pan, zoom and orbit around the 3D room to evaluate the placement.

The user was first asked to rate the plausibility of placement on a Likert Scale from 1-5. Following this, the user was asked if they wanted to move the object to a new location. If they answered "no", the user would progress to the next scene. If they answered "yes", the UI displayed transformation control handles (position axis arrows, rotation axis circles) to object position and orientation. After moving the object to the desired location, the user could save the placement and progress to the next scene. An IRB approval was maintained ahead of the experiment.

\section{Results}
\label{section:results}
\subsection{Ablation Studies}

\subsubsection{Position Features} In this experiment, we remove objects from test scenes taken from the Matterport3D dataset and replace it using various versions of our model in an ablation study. In Figure \ref{fig:abl_pos1}, we plot the cumulative distance between the ground truth position and the top position prediction, and in Figure \ref{fig:abl_pos2}, we plot the cumulative distance between the ground truth position and the nearest out of the the top 5 position predictions, using our full system and three ablated versions.

We find that the full SceneGen system predicts a placement most similar to ground truth than any of the ablated versions, followed by the models using AverageDist and RoomPosition features (AD+RP), and SurroundedBy and RoomPosition (S+RP). The predictions furthest from the ground truth are generated by only using the RoomPosition (RP) feature. These curves are consistent between the best and the closest of the top 5 predicted positions and indicate that each of our features for position prediction contributes to the accuracy of the final result.

In addition, when the top 5 predictions are considered, we see that each system we assessed is able to identify high probability zones closer to the ground truth. This is supported by the slope of the curves in Figure \ref{fig:abl_pos2}, evaluating the closest of the top 5 predictions, which rise much more sharply than in Figure \ref{fig:abl_pos1}, using the only best prediction. This difference provides support for the importance of predicting multiple locations instead of simply returning the highest-scored sampling location. A room can contain multiple plausible locations for a new object, so our system's most highly scored location will not necessarily be same as the ground truth's. For this reason, our system returns probabilities across sampled positions using a heat map to show multiple viable predictions for any placement query.

\begin{figure}
  \includegraphics[width=1\columnwidth]{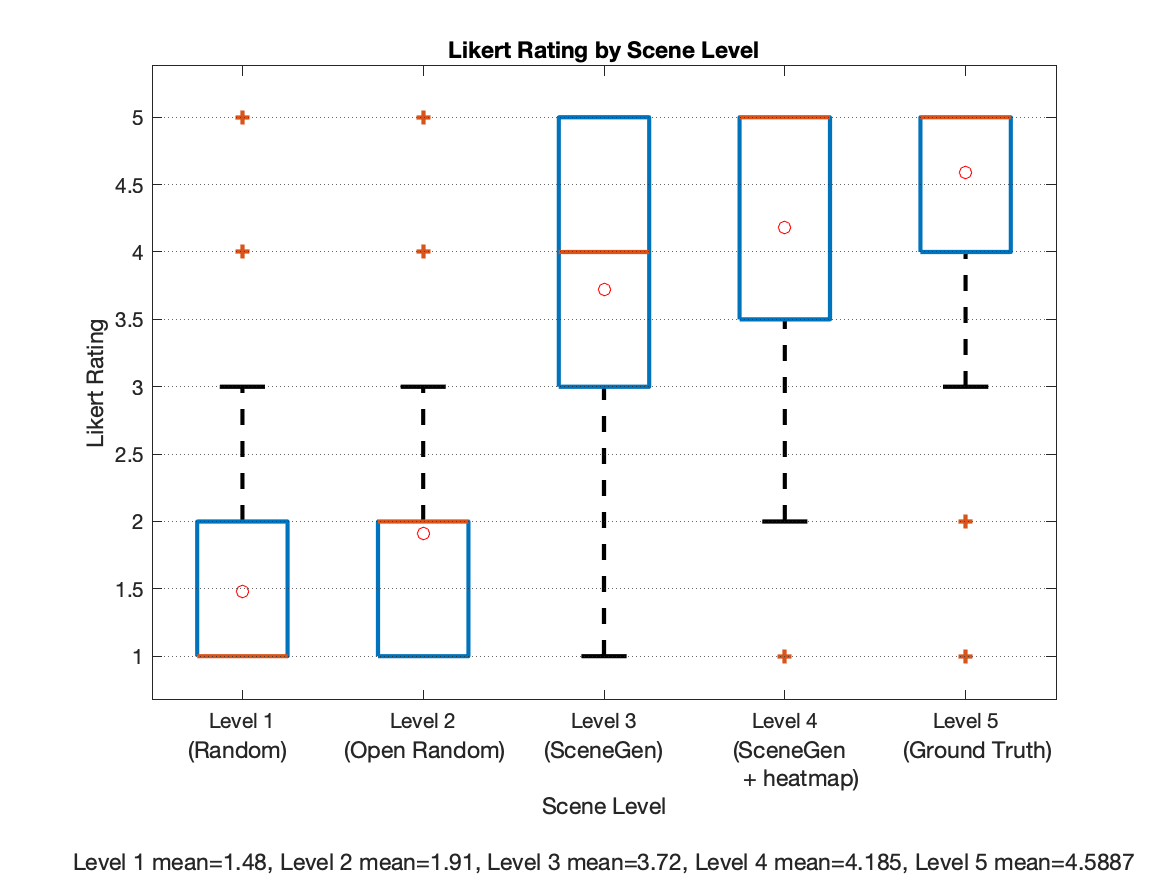}
  \caption{Users rate the plausibility of object placement in each room on the Likert Scale from 1 to 5. (1= Implausible/ Random, 3= Somewhat Plausible, 5 = Very Plausible). Scores are displayed in a box plot separated by the user study level.}~\label{fig:levels}
\end{figure}

\begin{figure}
  \includegraphics[width=1\columnwidth]{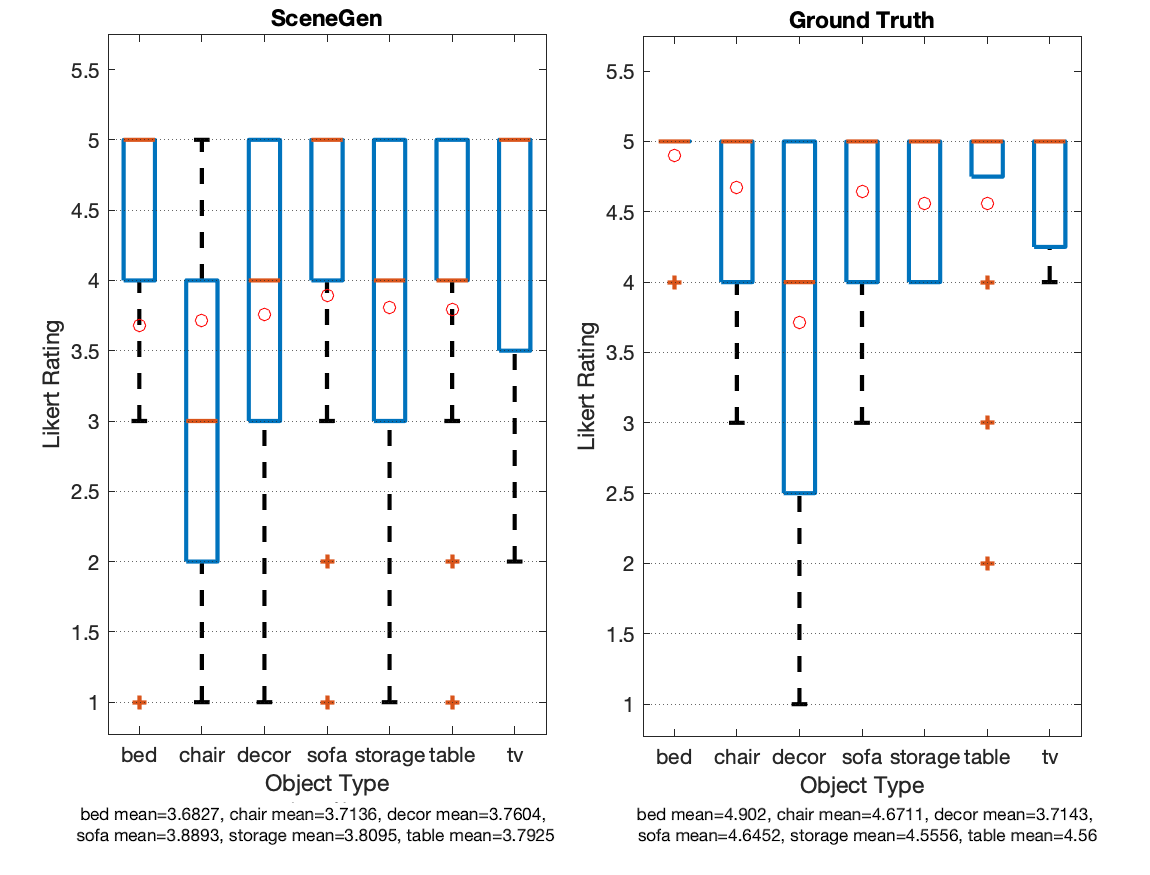}
  \caption{The plausibility score for each object category on the Likert Scale given by users is compared between SceneGen Levels (III, IV) and the ground truth, Level V. }~\label{fig:catboxplot}
\end{figure}

Table \ref{tab:table_ablation_pos}, shows the mean distance of the position prediction to ground truth position separated by object categories. We find that the object categories where the full SceneGen system outperforms its ablations are chairs, storage, and decor. For beds and TVs, SceneGen only produces the closest placements out of the system versions when considering the top five predictions. For pictures and tables, SceneGen's top prediction is closest to ground truth, and is only slightly further when comparing the nearest of the top 5 predictions.

\subsubsection{Orientation Features Results}\label{section:orientation_ablation_results} 

As with our position ablation studies, we assess the ability of various versions of our model to reorient assymmetric objects from test scenes. In Figure \ref{fig:abl_or}, we plot the angular distance between the ground truth orientation and the top orientation prediction, various versions of our system. The base model includes only Facing, (F), and is the lowest performing. We find that the system that also includes TowardsCenter and RoomPosition features performs best overall. We use this system (F+C+RP) in our implementation of SceneGen. The other four versions of our system perform similarly to each other overall.

Table \ref{tab:table_ablation_orien} shows the results of the orientation ablation study separated by object category. In this case, the system with Facing, TowardsCenter and RoomPosition features (F+C+RP) outperforms all other versions across on all categories except for TVs, where the system that includes Facing, TowardsCenter and NextTo (F+C+NT) produces the least deviation. In fact, all three of the systems that included either DirectionSimilarity or NextTo, predict the orientation of TVs more closely than the overall best performing system, but perform more poorly on other objects such as beds when compared with  systems without those features. This suggests that for other datasets, these features could be more effective in prediction orientations.

\subsection{User Study Results}

\begin{figure}
  \includegraphics[width=1\columnwidth]{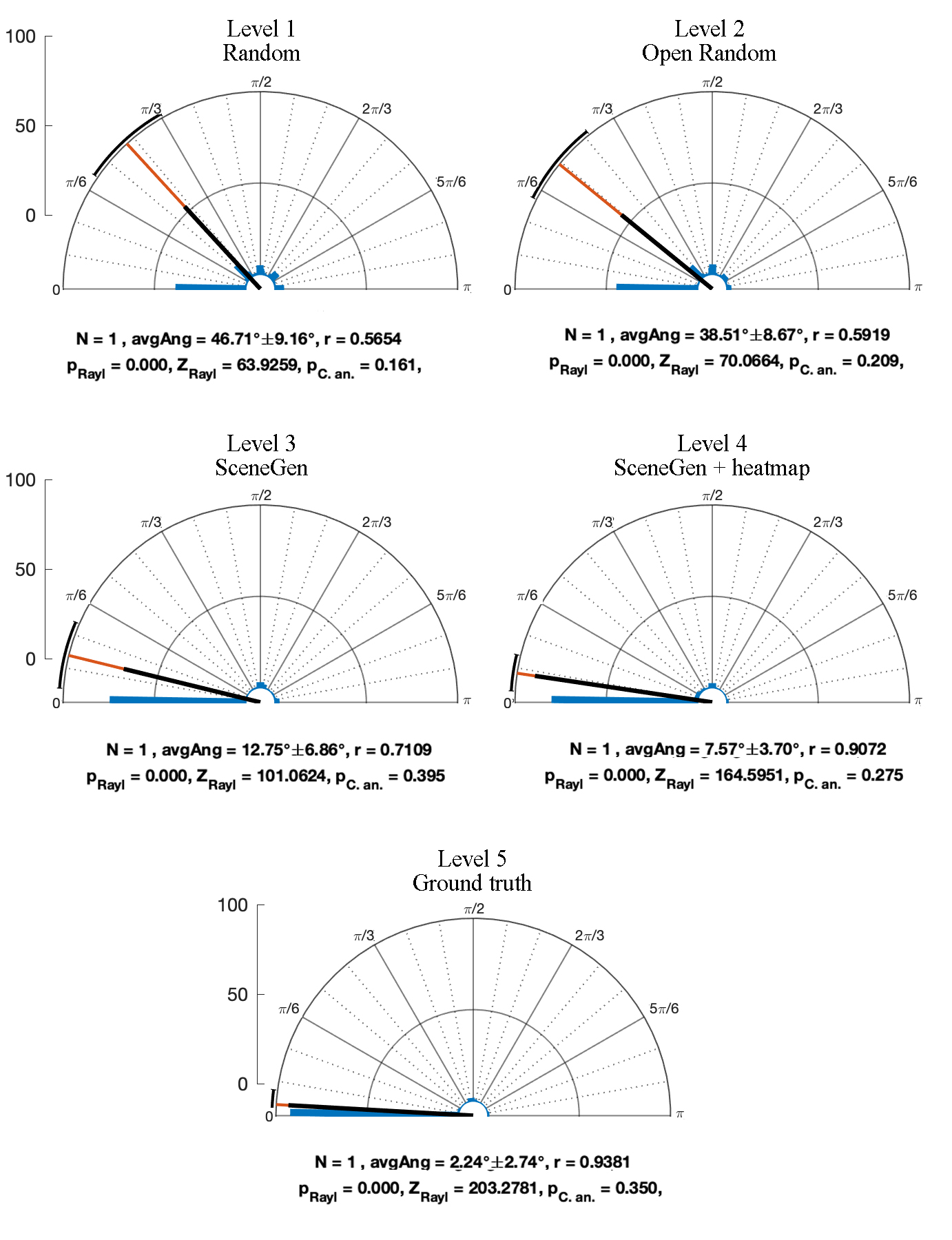}
  \caption{Radial histograms display distribution of how much a user rotated an object from its orientation in each level of the user study. Figure created using \cite{zittrell_2020}.}~\label{fig:radialgraph}
\end{figure}

\subsubsection{Plausibility of Placement Results}

We show the distributions of Likert ratings by level in Figure \ref{fig:levels}. We also run a one-way ANOVA test on the Likert ratings of initial placements, finding significant differences between all pairs of levels except for Levels IV and V. In other words, the ratings for Level IV’s representation of our prediction system are not significantly different from ground truth placements. Across multiple tests, we see that Level IV result means are significantly different from levels based on randomization, while Level III is only sometimes. As the Level IV presentation of the system can have multiple suggested initial placements, this difference between Levels III and IV could support our conjecture that accounting for multiple “right answer” placements improves the predictions.

\subsubsection{Position Prediction Results}
\begin{figure}
  \includegraphics[width=0.8\columnwidth]{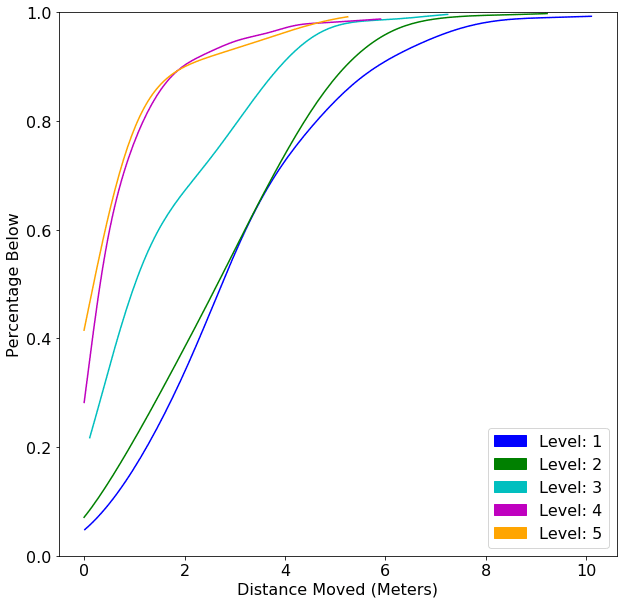}
  \caption{Cumulative density plot indicates the distance objects were moved from its placement in each level of the user study.}~\label{fig:userdistplot}
\end{figure}

We analyze how participants’ choices to adjust placement and amount moved varied across different scene levels. Results of this can be seen in Figure \ref{fig:userdistplot}. A one-way ANOVA test of the distance users moved objects from its placements found a significant difference ($p = 1.8622e^{38}$) between two groupings of levels: 1) Levels I and II (with higher means), and 2) Levels III, IV, and V (with lower means). This first group contains the levels with randomized initial placements, while this second group contains the levels that use our prediction system or the ground truth placement. The differentiation in groupings provides support for the plausibility of our system’s position predictions over random placements. 

\subsubsection{Orientation Prediction Results}
A one-way ANOVA test was performed on the overall change in object orientation from the participants’ manual adjustment, and found a significant difference ($p = 1.8112e^{16}$) between a different pair of level groupings: 1) Levels I, II, and III, and 2) Levels IV and V. In Figure \ref{fig:radialgraph}, we show the distributions of angular distance between the initial object orientation and the final user-chosen orientation, for each level. The levels IV and V have distributions are most concentrated at no rotation by the user. In Levels I and II, the users rotate objects more than half of the time, with an average rotation greater than $\frac{\pi}{6}$ radians. A vast majority of objects placed by Levels III, IV, V systems are not rotated by the user, lending support to the validity of our prediction system.

\section{Augmented Reality Application}
\label{section:ar}
To demonstrate a way to integrate our prediction system in action, we have implemented an augmented reality application that augments a scene using SceneGen. Users can overlay bounding boxes over the existing furniture to see the object bounds used in our predictions. On inserting a new object to the scene, the user can visualize a portability map to observe potential positions. Our Augmented Reality application consists of five main modules: (i) local semantic segmentation of the room; (ii) local Scene Graph generation (iii) heat map generation which is developed on an external server (iv) local data parsing and visualization; and finally (v) the user interface. We briefly discuss each of these modules below.

Semantic segmentation of the room can be done either manually or automatically, using integrated tools available on augmented reality devices. However, as not all current AR devices are equipped with depth-sensing capturing hardware, we use techniques previously introduced by \cite{saran2019augmented}, allowing the user themselves to manually generate and annotate semantic bounding boxes of objects of the target scene. The data acquired are then converted to our proposed spatial Scene Graph, resulting in an abstract representation of the scene. Both semantic segmentation and graph generation modules are performed locally on the AR devices, ensuring the privacy of the raw spatial data of the user.

Once the Scene Graph is generated, it is sent to a remote server where SceneGen engine can calculate positional and orientation augmentation probability maps for the target scene. The prediction probability maps for all objects are generated in this step. Such approach would allow faster computation time, since current AR devices come with limited computational and memory resources. The results are sent back to the local device, in which can be parsed and visualized using the Augmented Reality GUI.


The instantiation system can toggle between two modes: \textit{Manual} and \textit{SceneGen}. In Manual mode, the object is placed in front of the user, on the intersection of the camera front-facing vector direction with the floor. This would normally result in augmenting the object in the middle of the screen. While such conventional approach allows the user control the initial placement by determining the pose of the AR camera, in many cases additional movements are necessary to place the object in a plausible final location. In such cases, the user can then further move and rotate the objects to its desirable location. In SceneGen mode, the virtual object is augmented using the prediction of our system, resulting in faster and contextual placements.

\section{Discussion}

\subsection{Features and Predictions}
The Scene Graph we introduce in this paper is designed to capture spatial relationships between objects, object categories and the room. Overall, we have found that each of the relationships we have presented improves the model's ability to augment virtual objects in realistic placements in a scene. These relationships are important to understand the functional purposes of the space in addition to the individual objects.

In SceneGen, RoomPosition is used as a feature in predicting both orientation and position of objects. While this is a feature based solely on the position of the object, where it is in a room also has a strong impact on the function of the object and how it should be oriented. For example, a chair in a corner of the room is very likely to face towards the center of the room, while a chair in the middle of the room is more likely to face towards a table or a sofa. When analyzing our placement predictions probability maps and our user study results, we have observed that the best orientation is not the same at each position. This is not only affected by the nearby objects, but also by the sampled position within the room.

\begin{figure}
  \includegraphics[width=1\columnwidth]{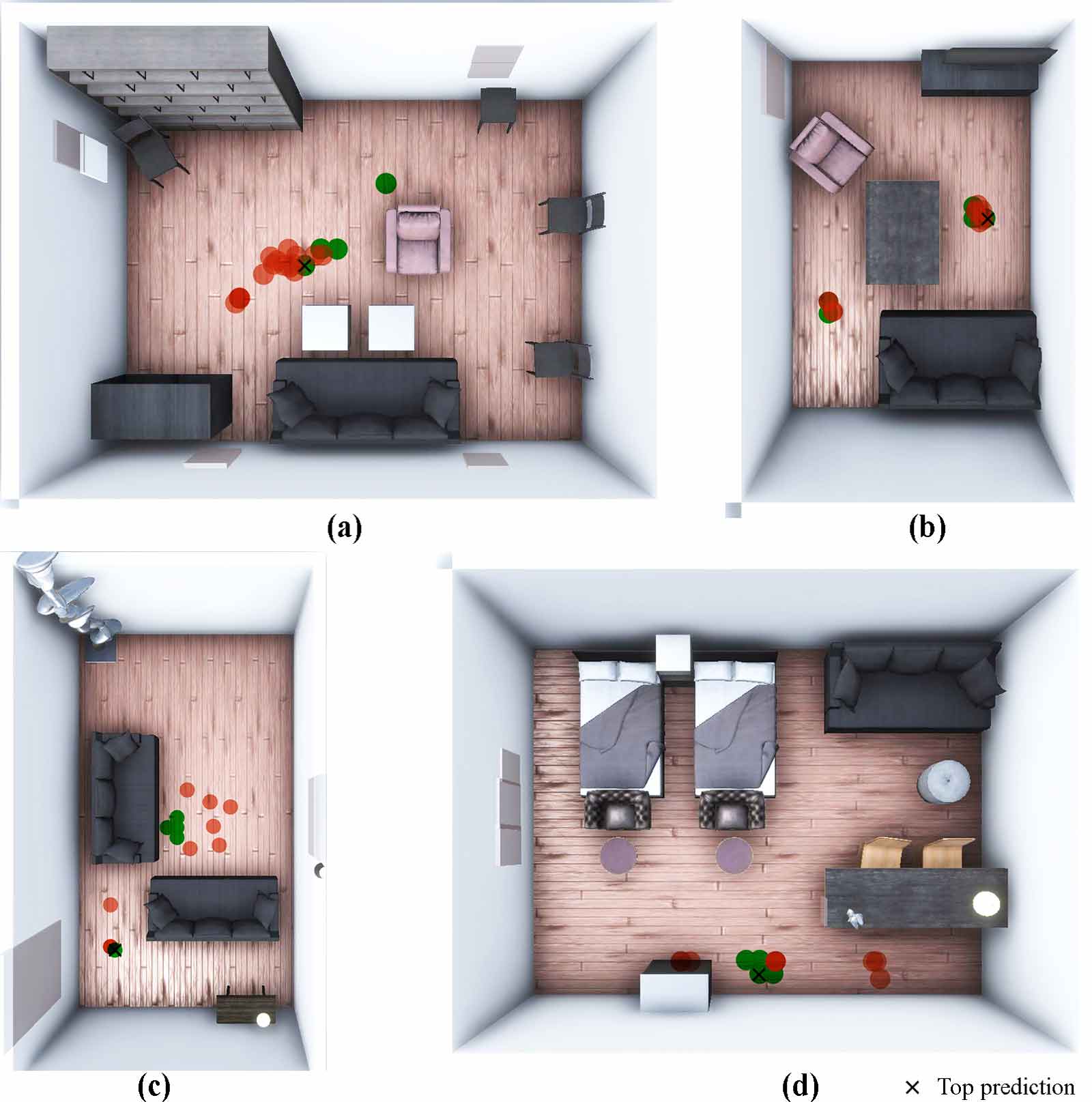}
  \caption{Top 5 highest probability positions for placing sofa (a,b), table (c) and TV (d) predicted by SceneGen (green) are compared to the user placements (red) showing that different users prefer different locations in a room and SceneGen also finds the clusters preferred by users to be highly probable. }~\label{fig:userpredictions}
\end{figure}

\subsection{Explicit Knowledge Model}
In our evaluation of SceneGen, we have found a number of benefits in using an explicit model to predict object placements. One benefit is that if we want to define a non-standard object to be placed in relation with standard objects by specifying your own relationship distributions, it is feasible with our system but would not be possible for implicit models. For example, in a collaborative virtual environment, where special markers are desired to be placed near each user, one could specify distributions for relationships such as NextTo chair and Facing table, without needing to train these from a dataset.

Another benefit is that explicit models can be easily examined directly to understand why objects are being placed where they are. For example, the Bed orientation feature distribution, based on the Matterport3D priors in Figure \ref{fig:relGraphs}, marginalized with respect to all other variables except TowardsCenter show that beds are nearly 5 times as likely to face the center of the room, while marginalizing features except position of the Storage show that a storage is found in a corner of a room 63\% of the time, along an edge 33\% of the time, and only in the middle of the room in 4\% of occurrences. 

\subsection{Dataset}
One important consideration in our choice of dataset is that we aim to learn spatial relationships for real world scenes. One can imagine idiosyncrasies of lived-in rooms, such as an office chair that is not always tucked into a desk but often left rotated away from it or a dining table pushed into a wall to create more space in a family room. Using personal living spaces, from the Matterport3D dataset, as our priors, we can capture these relationships that exist only in real world, lived-in scenes.  

One drawback of using the Matterport3D dataset is that it is not as large as some synthetic datasets. In our implementation, we group objects into broader groups to ensure adequate representation to ensure that all object categories are represented well enough to approximate the distribution of a large feature space. 

Another downside of using a real-world dataset is its accuracy in labeling where many human errors occur in this labour intensive process. Such mismatches are unlikely to happen in synthetic datasets as the geometry is already assigned in a digital format. To mitigate some of these concerns, we have developed a labeling application that allows us to determine the correct orientation of each objects, and also filter out rooms with corrupted scans and inaccurate labeling.

\begin{figure}
  \includegraphics[width=1\columnwidth]{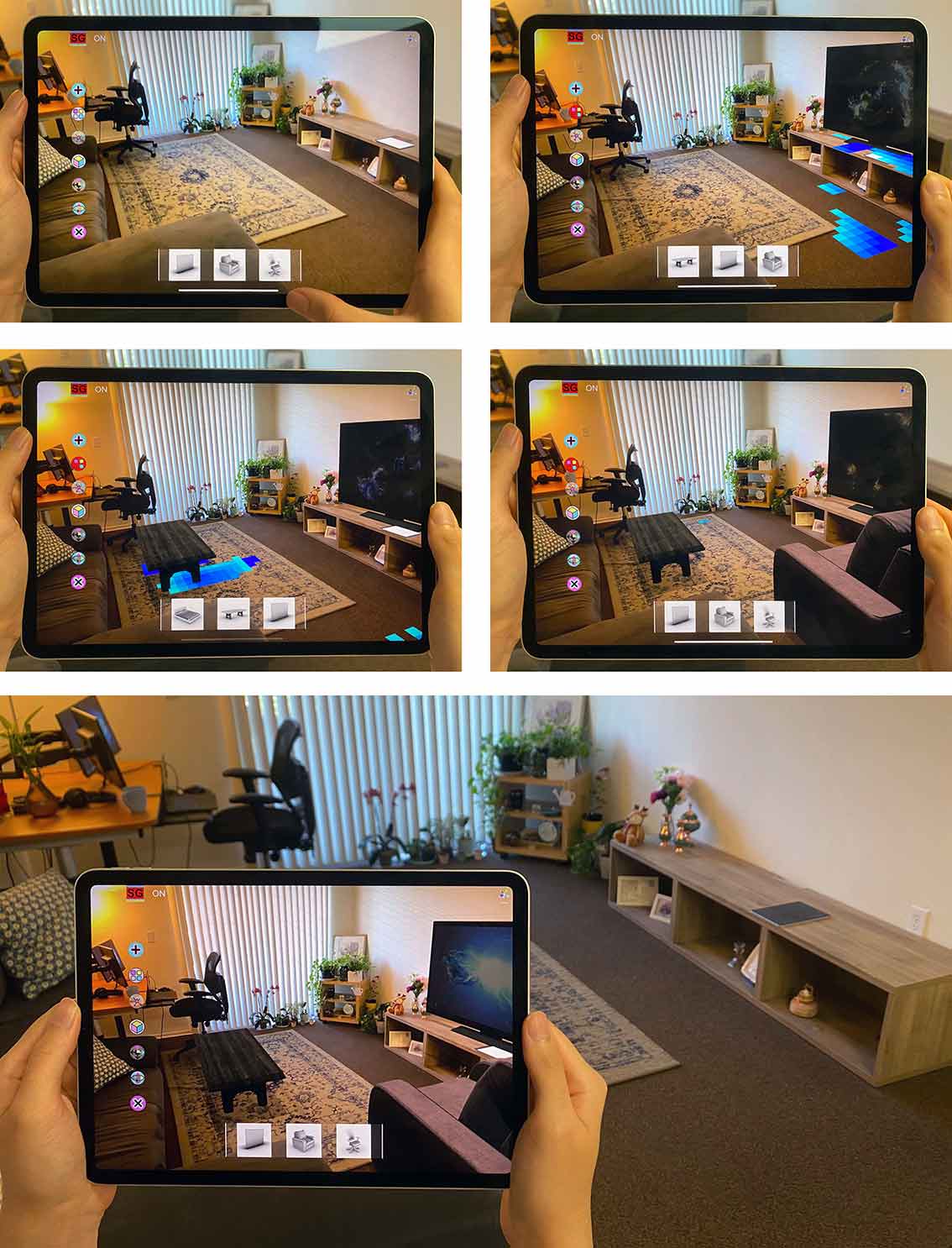}
  \caption{Augmented Reality application demonstrates how SceneGen can be used to add virtual objects to a scene. From the target scene (top-left), a TV (top-right), table (middle-left), and then a sofa (middle-right) are placed in the most probable poses. A probability map can be displayed indicating how likely each position is (top-left, middle-left). The AR application with virtual objects is compared to the original scene (bottom).  }~\label{fig:ArApp}
\end{figure}

\subsection{Subjectivity of Placements}
Where and how an object is placed in a scene is often very subjective and preferences can differ between users. This is demonstrated by the Likert scale plausibility ratings in Level V reference scenes in the user studies. Figures \ref{fig:levels} and \ref{fig:catboxplot} show that some users would only give scores of \textit{somewhat plausible} to scenes that are modelled from real world ground truth Matterport3D rooms. This supports providing a heat map of probabilities for each sampled placement, as alternate high probability positions may be more preferable to different users. Our results also indicate that most users prefer level IV scenes, with the heat map, compared to level III scenes, even though the placements use the same SceneGen models. This suggests that the inclusion of the heat map guides the users towards the system's placement and may help in convincing them of the viability and reasoning for such a choice. 

We also see that some users move objects to other high probability alternatives as seen in Figure \ref{fig:userpredictions}. This is a similar result to the position prediction experiment, which compares the ground truth position to the closest of SceneGen's top 5 predictions and shows that while the reference position may not always be the top prediction, it was often one of the top predictions. Moreover, results in Figure \ref{fig:catboxplot} show the subjectivity of an object placement is highly dependent on the size and type of object itself. In any room, there are very few natural places to put a bed. Hence the results for placing beds cluster in one or two high probability locations. Other objects such as decor are more likely to be subject to user preferences.

\section{Conclusion}
In this paper we introduce a framework to augment scenes with one or more virtual objects using an explicit generative model trained on spatial relationship priors. Scene Graphs from a dataset of scenes are aggregated into a Knowledge Model and used to train a probabilistic model. This explicit model allows for direct analysis of the learned priors and allows for users to input custom relationships to place non-standard objects alongside traditional objects.  SceneGen places the object in the highest probability pose and also offers alternate highly likely placements. 

We implement SceneGen using the Matterport3D, a dataset composed of 3D scans of lived-in rooms, in order to understand object relationships in a real world setting. The features that SceneGen extracts to build our Scene Graph are assessed through an ablation study, identifying how each feature contributes to our model's ability to predict realistic object placements. User Studies also demonstrate that SceneGen is able to augment scenes in a much more plausible way than system that places objects randomly or in open spaces. We also found that different users have their own preferences for where an object should be placed. Suggesting multiple high probability possibilities through a heat map allows users an intuitive visualization of the augmentation process.

There are of course, limitations to our work. While SceneGen is able to iteratively add objects to a scene, the resulting layout is highly dependent on the order in which objects are placed. Such approach does not consider all possible permutations of the possible arrangements. In addition, it can narrow down the open possible spaces for later objects, forcing placements that are far from optimal. Moreover, in scenarios where a large number of objects are to be augment, the current approach may not have the ability to \textit{fit} all the objects within the usable space, as initial placements are not aware of upcoming objects. Future work can comprise of incorporating  floorplanning methodologies with the current sampling mechanism allowing a robust search in the solution space, while addressing combinatorial arrangement. 

Moreover, SceneGen is a framework that naturally fits into spatial computing applications. We demonstrate this in a augmented reality application that augments a scene with a virtual object using SceneGen. Contextual scene augmentation can be useful in augmenting collaborative mixed reality environments or in other design applications, and using this framework allows for fast and realistic scene and content generation. We plan on improving our framework in providing the option to contextually augment non-standard objects by parameterizing topological relationships, a feature that would facilitate content creation for future spatial computing workflows.   

\begin{acks}
We acknowledge the generous support from the following research grants: FHL Vive Center for Enhanced Reality Seed Grant, a Siemens Berkeley Industrial Partnership Grant, ONR N00014-19-1-2066.
\end{acks}

\bibliographystyle{ACM-Reference-Format}
\bibliography{bibliography}

\appendix

\end{document}